\documentclass{aa}  
\usepackage{graphicx}
\usepackage{txfonts}
\usepackage{amsmath}
\usepackage{amssymb}
\usepackage{listings}
\usepackage{xcolor}
\usepackage[colorlinks=true, linkcolor=blue, citecolor=blue, filecolor=blue, urlcolor=blue]{hyperref}
\begin{document}

\title{A Method for Determining the Locations and Configurations of Magnetic Reconnection within 3D Turbulent Plasmas}

\author{Yulei Wang
        \inst{1,2}
        \and
        Xin Cheng\inst{1,2}
        \and
        Yang Guo\inst{1,2}
        \and
        Jinhan Guo\inst{1,2,3}
        \and
        Mingde Ding\inst{1,2}
}

\institute{School of Astronomy and Space Science, Nanjing University, 
           Nanjing 210023, People's Republic of China \\
           \email{wyulei@nju.edu.cn,xincheng@nju.edu.cn}
         \and
           Key Laboratory for Modern Astronomy and Astrophysics (Nanjing University), 
           Ministry of Education, Nanjing 210023, People's Republic of China
         \and
           Centre for Mathematical Plasma Astrophysics, Department of Mathematics, KU Leuven, Celestijnenlaan 200B, B-3001 Leuven, Belgium
}

\abstract
{
   Three-dimensional (3D) reconnection is an important mechanism for efficiently releasing energy during astrophysical eruptive events, which is difficult to be quantitatively analyzed especially within turbulent plasmas.
}
{
   In this paper, an efficient method for identifying locations and configurations of 3D reconnection from MHD data is developed.
}
{
   This method analyzes the local nonideal electric field and magnetic structure at an arbitrary position. 
   As only performing algebraical manipulations on the discrete field data and avoiding computationally expensive operations like field-line tracing and root-finding, this method naturally possesses high efficiency.
   To validate this method, we apply it to the 3D data from a high-resolution simulation of a Harris-sheet reconnection and a data-driven simulation of a coronal flux rope eruption.
}
{
   It is shown that this method can precisely identify the local structures of discrete magnetic field.
   Through the information of nonideal electric field and the geometric attributes of magnetic field, the local structures of reconnection sites can be effectively and comprehensively determined.
   For fine turbulent processes, both qualitative pictures and quantitative statistical properties of small-scale reconnection structures can be obtained.
   For large-scale solar simulations, macro-scale magnetic structures such as flux ropes and eruption current sheets can also be recognized.
}
{
   We develop a powerful method to analyze multi-scale structures of 3D reconnection.
   It can be applied not only in MHD simulations but also in kinetic simulations, plasma experiments, and \emph{in-situ} observations.
}

\keywords{3D magnetic reconnection -- turbulence -- data analysis}

\titlerunning{A Method for Analyzing Magnetic Reconnection within 3D Turbulent Plasmas}
\authorrunning{Wang, Yulei, et al.}
\maketitle
%

\section{Introduction\label{sec:Intro}}

Magnetic reconnection is believed to be the driving mechanism of various astrophysical eruptive phenomena, during which magnetic energy is rapidly released to heat plasma and accelerate charged particles.
The classic theory of magnetic reconnection is built on two-dimensional (2D) steady models, but in realistic physical environments, the three-dimensional (3D) effects cannot be neglected \citep{Priest2000}.
Far more complex than 2D scenarios, 3D reconnection is hard to be investigated analytically, especially for unsteady reconnection processes coupled with significantly multi-scale and nonlinear characteristics.
Therefore, numerical simulations play vital roles in studies of 3D reconnection \citep{Ji2022}.
With the recent development of large-scale high-performance computers, high-resolution simulations of both kinetic and MHD scales have been performed, which proves the formation of self-sustained turbulence induced by 3D reconnection \citep[see][]{Huang2016,Kowal2020,Zhang2021,Comisso2022,Dong2022,Wang2023a}.
Within the turbulent regions, the large-scale structures keep cascading toward smaller ones following a power-law spectrum, resulting in complicated patterns and chaotic magnetic structures. 
Though high-resolution simulations provide an unprecedented opportunity for understanding the fine processes of 3D reconnection, how to efficiently locate the reconnection sites and analyze their properties from the massive discrete data has become a new challenge \citep{Vlahos2023}.

A typical location of 3D reconnection is at a magnetic null point, where the magnetic strength vanishes.
In the adjacent region of a null point, magnetic field forms a fan-spine structure, the local structures of which can be evaluated by the magnetic Jacobian matrix $\mathbf{D}_B=\partial_jB^i$, where $\partial_j\equiv \partial/\partial x_j$ and $i,j=1,2,3$ \citep[see][]{Lau1990,Parnell1996}.
\citet{Haynes2007} developed a trilinear method for locating 3D null points in discrete numerical data, which has been widely applied and can also be directly used in turbulent cases. 
Methods for analyzing null points from satellite data have also been developed \citep[see][]{Fu2015,Olshevsky2020, Zhang2023}.

However, 3D reconnection can happen without the presence of null points, thus being more difficult to be analyzed.
\citet{Schindler1988} and \citet{Hesse1988} proposed the general magnetic reconnection theory and obtained the condition for the ``global'' effect of reconnection with finite magnetic field, namely,
\begin{equation}
   \int_{P_1}^{P_2} E_{\parallel}\mathrm{d}s\neq 0\,,\label{eq:def_3DRec}
\end{equation}
where $E_{\parallel}$ is the electric field strength parallel with magnetic field $\bf B$, the integration is taken along a magnetic field line, and the ``global'' effect refers to the two plasma elements in the ideal region, $P_1$ and $P_2$, can ``feel'' the reconnection and will move to different field lines later.

3D reconnection can take place at locations where strong local current concentrates and various types of 3D reconnection have been discovered such as the torsional/shearing reconnection within the fan/spine near a null point, the separator reconnection, the hyperbolic-flux-tube (HFT) reconnection, and the braid reconnection \citep[see][]{Pontin2022}.  
Meanwhile, the local magnetic structures at the reconnection region determine the properties of 3D reconnection.
\citet{Priest1989} proposed the singular line (SL) reconnection which forms X-type magnetic structures on the plane perpendicular to the local magnetic field.
\citet{Wilmot-Smith2007} provided an example of 3D reconnection at the center of a flux tube with footpoints rotating toward different directions.
\citet{Parnell2010} found the O-type reconnection magnetic structures even along a separator connecting two null points in their simulations. 

To analyze locations of 3D reconnection from the discrete magnetic field data obtained in simulations, observations, and experiments, both global and local methods have been developed.
The global methods, relying on the field-line tracing technique, are closely related to the 3D reconnection theory but mainly suitable for analyzing large-scale laminar magnetic fields.
For example, the quasi-separatrix layer (QSL) method, first proposed by \cite{Priest1995} and later improved by \cite{Titov2002} and \cite{Titov2007}, has been widely applied to study coronal magnetic field \citep{Demoulin1996,Demoulin1996a,Demoulin1997,Aulanier2006,Titov2009,Li2021a,Zhong2021,Li2022,Guo2023a}.
QSLs are locations where the connectivity of magnetic field, evaluated by the squashing factor $Q$, changes significantly.
Though $Q$-factor is not a direct indicator of reconnection \citep[][]{Reid2020}, it is useful for investigating magnetic topologies and exhibiting locations where 3D reconnection might happen.
The calculation of $Q$-factor for discrete data relies on field-line tracing, which can consume significant computation resources without appropriate optimizations.
Recently, the efficiency of the QSL method has been improved via the hardware accelerations of GPUs \citep[]{Zhang2022}.
As another algorithm relying on field-line tracing, the method developed by \cite{Haynes2010} can detect the global laminar magnetic topology (magnetic skeleton) in numerical results, including null points, spines, separatrix surfaces, and separators.
By using this method, \cite{Parnell2010a} statistically proved the importance of separator reconnection during the interaction between emerging and overlying flux. 
\cite{Komar2013} proposed a similar method for tracing the separators within discrete data based on the pre-located null points.

Different from global methods, the local analysis focuses on the differential structures depending only on the adjacent region of a position and thus naturally possesses high efficiency.
More importantly, it also provides statistical laws that are vital for understanding the physics of turbulent reconnection.
As a straightforward method, the distributions of $E_{\parallel}$ and $J_{\parallel}$ can reflect the sites of 3D reconnection according to the definition Eq.\,(\ref{eq:def_3DRec}), which has been used in the visualization of turbulent reconnection simulations \citep[see][]{Huang2016,Isliker2019,Dong2022}.

For 3D systems imposed with a strong uniform background magnetic field, the methods of locating reconnection structures can be simplified as 2D methods.
For instance, \cite{Zhdankin2013} proposed an algorithm for identifying the geometrical and physical properties of current sheets.
\cite{Li2021b} developed the magnetic flux transport method which was recently applied to a kinetic plasma turbulence simulation by \cite{Li2023b}.

For general 3D cases without a strong guide field, local reference frames are necessary to analyze small-scale structures.
\cite{Parnell2010} set the direction of a pre-determined separator as the normal direction of the projection plane, on which the 2D projected magnetic field is classified. 
\cite{Kowal2020} defined a local frame based on the shear tensor of magnetic field, which is further used to determine the properties of current sheets.
Recently, \cite{Lapenta2021} developed a reconnection-site identification method via the electric drift speed, which performs well for data of kinetic simulations.

Methods independent with local frames have also been developed for the identification of turbulent reconnection structures \citep[see][]{Vlahos2023}, for instance, the Phase Coherence Index method \citep{Hada2003}, the Partial Variance of Increments (PVI) method \citep{Greco2017}, and the fractal dimension method by box-counting \citep{Isliker2019}.
Although these methods can produce useful statistical results, they cannot analyze the magnetic structures at reconnection sites.

In this paper, we introduce an efficient method for locating small-scale reconnection structures in 3D discrete data involving turbulent eddies.
Both the theoretical basis and the numerical algorithm of this method are presented in detail.
Through two typical benchmarks, it is well validated and presents a promise in qualitative and quantitative analysis of 3D reconnection simulations containing multi-scale structures.

The following parts of this paper are organized as follows.
In Sect.\,\ref{sec:theory}, we exhibit the theoretical basis of our method.
The numerical strategy is presented in Sect.\,\ref{sec:method}.
We provide a \textsf{Matlab} implementation of this method and introduce its usage in Sect.\,\ref{sec:LoRD}.
This method is benchmarked by two typical simulations which are detailed in Sects.\,\ref{sec:test1} and \ref{sec:test2}.
The results of this paper are summarized and discussed in Sect.\,\ref{sec:conclude}.

\section{Theory \label{sec:theory}}

\subsection{The General Magnetic Reconnection Theory\label{ssec:GMR}}

We first briefly summarize the main concepts of the general magnetic reconnection theory developed by \citet{Hesse1988}.

Nonvanishing magnetic field can be (at least locally) expressed by the Euler potentials as $\bf{B}=\nabla\alpha\times\nabla\beta$.
The three vectors $\nabla\alpha$, $\nabla\beta$, and $\hat{\mathbf{b}}=\mathbf{B}/\left|\mathbf{B}\right|$ compose a local frame system $\left(\alpha,\beta,s\right)$, where $s$ denotes the arc coordinate along a field line.
An $\left(\alpha,\beta\right)$-pair corresponds to a field line.
$\nabla\alpha$ and $\nabla\beta$ span the local plane normal to $\bf B$ but, generally speaking, they are not orthogonal or normalized. 

If setting the vector potential as $\bf A=\alpha\nabla\beta$ satisfying the gauge $\bf A\cdot B=0$, Ohm's law $\bf E=-v\times B+N$ and Faraday's law produce the evolution rule of the Euler potentials as
\begin{subequations}\label{eq:GMR}
\begin{align}
   \dot{\alpha} & = -\frac{\partial\psi}{\partial\beta}-N^{\beta}\,,\\
   \dot{\beta} & = \frac{\partial\psi}{\partial\alpha}-N^{\alpha}\,,
\end{align}
\end{subequations}
where $\bf N$ is the general nonideal term, $\psi=\phi+\alpha\partial\beta/\partial t$, $\phi$ is the electric scalar potential, and the dot operator denotes the total derivative of time $\mathrm{d}/\mathrm{d}t$.
Notice that $\partial\psi/\partial s=-N^s=-E_\parallel$.

\citet{Hesse1988} generalized the definition of reconnection with finite magnetic field by the violation of line conservation.
Supposing two plasma elements $P_1$ and $P_2$ outside the nonideal region are connected by a field line, they will later move to two different field lines if and only if $\psi$ in Eq.\,\ref{eq:GMR} initially takes different values at $P_1$ and $P_2$, which directly results in the condition Eq.\,\ref{eq:def_3DRec}.
Because $\bf N=0$ at $P_1$ and $P_2$, the global effect is independent with the perpendicular components of $\bf N$, namely, $N^\alpha$ and $N^\beta$.

Different from 2D reconnection taking place at an X-point, 3D reconnection happens within a finite nonideal volume \citep[][]{Priest2003}, in which there exist ``special'' field lines.
After defining a quasi-potential $\Xi\left(\alpha,\beta\right)\equiv -\int_{P_1}^{P_2}E_\parallel\left(\alpha,\beta,s\right)\mathrm{d}s$, the global differences between the evolutions of Euler potentials at $P_1$ and $P_2$ can be expressed by \citep[][]{Hesse2005}
\begin{subequations}\label{eq:GMRDiff}
\begin{align}
   \left.\dot{\alpha}\right|_{P_2}-\left.\dot{\alpha}\right|_{P_1} & = -\frac{\partial\Xi}{\partial\beta}\,,\\
   \left.\dot{\beta}\right|_{P_2}-\left.\dot{\beta}\right|_{P_1} & = \frac{\partial\Xi}{\partial\alpha}\,.
\end{align}
\end{subequations}
According to this equation, if $\partial\Xi/\partial\alpha=\partial\Xi/\partial\beta=0$ ($\Xi$ has an extremal value in the $\alpha$-$\beta$ space), $P_1$ and $P_2$ will remain on the same field line and thus cannot feel the global reconnection effect.
Notice that $\partial\Xi/\partial\alpha=\partial\Xi/\partial\beta=0$ is equivalent to $\nabla_\perp\Xi=0$, where $\nabla_\perp$ denotes the gradient perpendicular to $\bf B$.
Moreover, \citet{Hesse2005} generalized the definition of reconnection rate to cases without separators or separatrices and proved that the extremal value of $\Xi$ equals the reconnection rate.
The extremal-$\Xi$ lines can be treated as generalizations of magnetic neutral lines or separators.
\citet{Wyper2015} later generalized this theory to complex nonideal regions with multiple peaks of $\Xi$.

\subsection{Local Analysis of 3D Reconnection\label{ssec:GMRLocal}}

The theory of global reconnection is well-defined but is not convenient to directly apply for analyzing discrete magnetic data, especially for cases with strong turbulence.
As revealed by various high-resolution simulations, the 3D reconnection regions are composed of multi-scale current sheets that are far more complex than theoretical models \citep[see][]{Dong2022,Wang2023a,Ye2023a}, which significantly enhances the difficulties in selecting appropriate start/end positions for field-line mappings.
Complete samplings of field lines within the entire system might produce reliable results, which, however, can hardly be operated as limited by computational resources and numerical integration errors.
Moreover, besides global information, local reconnection effects and structures are also valuable for studying the mechanism of 3D reconnection.
Therefore, we should seek local parameters of reconnection.

According to the condition Eq.\,\ref{eq:def_3DRec}, a \emph{necessary} condition for a field line to undergo the global reconnection is that it passes nonideal regions with $E_\parallel\neq 0$. 
In other words, we can use the distribution of $E_\parallel$ to exhibit regions where global reconnection effects can happen.
Provide that only Joule dissipation is considered and the resistivity $\eta$ is uniform, $E_\parallel$ can be replaced by $J_\parallel$ since $E_\parallel=\mathbf{N\cdot \hat{b}}=\eta\mathbf{J\cdot\hat{b}}=\eta J_\parallel$. 

Within the $E_\parallel\neq 0$ regions, we can further locate field lines with extremal values of $\Xi$.
A \emph{sufficient} condition for a field line to have an extremal $\Xi$ is that all the $E_\parallel\neq 0$ positions threaded by this field line satisfy $\nabla_\perp E_\parallel=0$, which can be directly proven by
\begin{equation}
   0=\int_{P_1}^{P_2}\nabla_\perp E_\parallel\mathrm{d}s=\nabla_\perp\int_{P_1}^{P_2}E_\parallel\mathrm{d}s=\nabla_\perp\Xi\,.
\end{equation}
In other words, in nonideal regions, locations with $\nabla_\perp E_\parallel=0$ approximately reveal the distributions of extremal-$\Xi$ lines, though the trajectories of extremal-$\Xi$ lines are not necessary to always pass through such locations.
$E_\parallel$ and $\nabla_\perp E_\parallel$ are two useful and easy-to-calculate local parameters that can provide implications about the locations of global reconnection effects, though they are not equivalent to the original integral definitions in Eqs.\,\ref{eq:def_3DRec} and \ref{eq:GMRDiff}.

The local reconnection effects, felt by plasma elements inside nonideal regions, correspond to the local violation of line conservation, which are blocked out by the global reconnection theory but still worth investigating.
The equation governing the local effects can be obtained by performing $\partial/\partial s$ on both sides of Eq.\,\ref{eq:GMR}, namely,
\begin{subequations}\label{eq:GMRLocal}
\begin{align}
   \frac{\partial\dot{\alpha}}{\partial s} & = \frac{\partial E_\parallel}{\partial\beta}-\frac{\partial N^{\beta}}{\partial s}\,,\\
   \frac{\partial\dot{\beta}}{\partial s} & = -\frac{\partial E_\parallel}{\partial\alpha}-\frac{\partial N^{\alpha}}{\partial s}\,,
\end{align}
\end{subequations}
which implies that the change of $\left(\dot{\alpha},\dot{\beta}\right)$ along a magnetic field line is governed by two terms, namely, the perpendicular gradient of $E_\parallel$ and the parallel gradient of $\mathbf{N}_\perp=\left(N^\alpha,N^\beta\right)$.
It can be proved that the line conservation ($\partial\dot{\alpha}/\partial s=\partial\dot{\beta}/\partial s=0$) is equivalent to $\bf B\times\left(\nabla\times N\right)=0$, while the condition of flux conservation is $\bf \nabla\times N=0$ \citep[][]{Hesse1988}.

Furthermore, it is helpful to evaluate the importance of the $E_\parallel$ and $\bf N_\perp$ terms in Eq.\,\ref{eq:GMRLocal}.
For instance, at locations with $\nabla_\perp E_\parallel=0$, the violation of line conservation is fully determined by the $\bf N_\perp$ term.
Moreover, suppose that the effects of $\bf N_\perp$ term are ignorable, the local effects would be dominated by the $E_\parallel$ term, which means that analysis of $E_\parallel$ can reveal the properties of not only global but also local effects of reconnection.
However, because a quantitative comparison of the two terms relies on a local flux coordinate frame $\left(\nabla\alpha,\nabla\beta\right)$, which is difficult to establish for arbitrary magnetic fields.
In Appendix \ref{asec:localeffects}, we attempt to discuss this problem and propose a statistical method to compare the effects of the two terms.
\\
\subsection{Magnetic Structures at Reconnection Sites\label{ssec:topology}}

As the magnetic null points have been well studied \citep[see][]{Parnell1996}, here we analyze the local magnetic structures near reconnection locations with finite magnetic field.
To begin with, we define a local frame at the origin $\mathbf{r}_0$ by three orthogonal unit vectors satisfying $\hat{\mathbf{e}}_3=\hat{\mathbf{b}}\left(\mathbf{r}_0\right)$, $\hat{\mathbf{e}}_2\perp\hat{\mathbf{e}}_3$, and $\hat{\mathbf{e}}_1=\hat{\mathbf{e}}_2\times\hat{\mathbf{e}}_3$.
In this frame, the magnetic field at $\mathrm{d}\mathbf{r}=\mathbf{r}-\mathbf{r}_0$ can be approximated by
\begin{equation}
   \mathbf{B}\left(\mathrm{d}\mathbf{r}\right) = B_0\hat{\mathbf{e}}_3+\mathrm{d}\mathbf{r}\cdot\nabla\mathbf{B}\left(\mathbf{r}_0\right)\,,\label{eq:B3D}
\end{equation}
where $B_0 = \left|\mathbf{B}\left(\mathbf{r}_0\right)\right|>0$ is the magnetic strength at $\mathbf{r}_0$ and $\left|\mathrm{d}\mathbf{r}\right|$ should be small enough to guarantee the validation of linear-field assumption.
$\hat{\bf{e}}_1$ and $\hat{\bf{e}}_2$ span the same normal plane $\mathrm{d}r_3=0$ as established by $\nabla\alpha$ and $\nabla\beta$.

The second term in Eq.\,\ref{eq:B3D} can be further written as two terms,
\begin{equation}
   \mathrm{d}\mathbf{r}\cdot\nabla\mathbf{B}\left(\mathbf{r}_0\right) = \left[\left(
      \begin{array}{c}
         \mathrm{d}r_1\\
         \mathrm{d}r_2\\
         0 
      \end{array}
   \right)+ \left(
      \begin{array}{c}
         0\\
         0\\
         \mathrm{d}r_3 
      \end{array}
   \right)\right]\cdot\nabla\mathbf{B}
    = \left(\begin{array}{c}
      B_\perp^1\\
      B_\perp^2\\
      B_\parallel
   \end{array}\right) + \tilde{\bf B}\,,\label{eq:B3DSplit}
\end{equation}
where $B_\perp^1$ and $B_\perp^2$ are two components of the perpendicular magnetic field defined by
\begin{align}
   \mathbf{B}_\perp & = \mathbf{M}\cdot\mathbf{R}\,,\label{eq:B2D}\\
   \mathbf{M} & = \left(\begin{array}{cc}
      \partial B^1/\partial R_1 & \partial B^1/\partial R_2\\
      \partial B^2/\partial R_1 & \partial B^2/\partial R_2
   \end{array}\right) \equiv\left(\begin{array}{cc}
      M_{11} & M_{12}\\
      M_{21} & M_{22}
   \end{array}\right)\,,\label{eq:M2D}\\
   \mathbf{R} & =\left(\begin{array}{c}
      \mathrm{d}r_1\\
      \mathrm{d}r_2\end{array}\right) \equiv \left(\begin{array}{c}
      R_1\\
      R_2
   \end{array}\right)\,,
\end{align}
and $B_\parallel=R_1\left(\partial B^3/\partial R_1\right)+R_2\left(\partial B^3/\partial R_2\right)$ is the parallel component. 
$\mathbf{B}_\perp$ and $B_\parallel$ compose the magnetic field lying in the $\mathrm{d}r_3=0$ plane, while $\tilde{\bf B}\equiv\mathrm{d}r_3\left(\partial\mathbf{B}/\partial\mathrm{d}r_3\right)$ represents the increments at locations of $\mathrm{d}r_3\neq0$.
Hereafter, we analyze the local in-plane magnetic structures on the $\mathrm{d}r_3=0$ plane following the ideas of the SL reconnection by \citet{Priest1989} and the O-type separator reconnection by \citet{Parnell2010}.
Therefore, $\tilde{\bf B}$ and $B_\parallel$ is not investigated in detail.
Suppose that the magnitudes of $\left|B_\parallel\right|$ and $\left|\tilde{\bf B}\right|$ are small enough compared with $\left|\mathbf{B}_\perp\right|$, the magnetic structure within a finite volume near $\mathbf{r}_0$ can be well approximated by $\mathbf{B}_\perp$.

Because $\mathbf{B}_\perp$ is not a real magnetic field satisfying the divergence-free condition, the trace of $\mathbf{M}$, evaluated by $tr\left(\mathbf{M}\right)=M_{11}+M_{22}=-\partial B^3/\partial \mathrm{d}r_3$, can be nonzero.
Meanwhile, one should notice that $\mathbf{B}_\perp$ vanishes at the frame origin ($\mathbf{R}=\left(0,0\right)^T$), where the superscript ``$T$'' denotes the transpose operation.
In other words, the definition of the local reference frame makes an arbitrary position $\mathbf{r}_0$ a 2D null point of $\mathbf{B}_\perp$.
Therefore, its 2D local structures can be fully determined by $\bf M$, which simplifies the theoretical analysis.
Moreover, for discrete fields, the structures of $\mathbf{B}_\perp$ at an arbitrary grid can be directly obtained via $\bf M$ and the computational costs of manipulations like interpolation and root-finding for searching null points between grids can be saved. 

Inspired by the method by \cite{Parnell1996}, we rewrite $\mathbf{M}$ as
\begin{equation}
   \mathbf{M} = \left(\begin{array}{cc}
      M_{11} & \left(q-J_3\right)/2\\
      \left(q+J_3\right)/2    & M_{22}
   \end{array}\right)\,,\label{eq:M2DM}
\end{equation}
where $q=M_{12}+M_{21}$ and $J_3=J_\parallel=M_{21}-M_{12}$.
Defining a threshold current $J_{\mathrm{thres}}=\sqrt{\left(M_{11}-M_{22}\right)^2+q^2}$, the discriminant of the eigenvalue quadratic $\left|\lambda\mathbf{I}-\mathbf{M}\right|=0$, becomes $D\left(\mathbf{M}\right)=4Det\left(\bf M\right)-tr\left(\bf M\right)^2=J_3^2-J_{\mathrm{thres}}^2$, where $Det\left(\bf M\right)$ is the determinant of $\bf M$.
The eigenvalues are thus $\lambda=\left(tr\left(\bf M\right)\pm\sqrt{-D\left(\bf M\right)}\right)/2$.
If $\left|J_3\right|<J_{\mathrm{thres}}$ (i.e. $D\left(\mathbf{M}\right)<0$), the eigenvalues of $\mathbf{M}$ are distinct real numbers; if $\left|J_3\right|>J_{\mathrm{thres}}$ (i.e. $D\left(\mathbf{M}\right)>0$), the eigenvalues of $\mathbf{M}$ are two conjugate complex numbers; if $\left|J_3\right|=J_{\mathrm{thres}}$ (i.e. $D\left(\mathbf{M}\right)=0$), two eigenvalues are repeated real roots.

The local structures of $\mathbf{B}_\perp$ can be classified into nine types based on the values of $tr\left(\bf M\right)$, $D\left(\bf M\right)$, and $\lambda$ (see Table\,\ref{tab:types}).
Type 1 has a local X-type structure, being the 3D generalization of 2D X-points and the location of SL reconnection if $E_\parallel\neq 0$ (see Fig.\,\ref{fig:diagram}a).
Types 2 and 3 are 3D O-types but possess radial components from the trace of $\bf M$.
For different signs of $tr\left(\bf M\right)$, field lines of types 2 and 3 show different directions toward $\mathbf{r}_0$ (see Fig.\,\ref{fig:diagram}b, c).
Types 4 and 5 are respectively source and sink structures dominated by the trace of $\bf M$ (see Fig.\,\ref{fig:diagram}d, e).
Type 6 has one zero eigenvalue, implying the existence of a neutral line (see Fig.\,\ref{fig:diagram}f).
The other types with zero traces are of 2D configurations, which are unstable and very rare in 3D evolutions \citep{Priest2000}. 
Types 7 and 8 are 2D X and O points, while type 9 are anti-parallel lines \citep{Parnell1996}.

\begin{table}
\caption{Types of $\mathbf{B}_\perp$.
The star marker ``$*$'' denotes complex conjugate. $\mathbb{R}$ and $\mathbb{C}$ are the sets of real and complex numbers, respectively.}
\label{tab:types}
\centering
\begin{tabular}{c c c c c}
\hline\hline 
Type & $tr\left(\bf M\right)$ & $D\left(\bf M\right)$ & $\lambda$ & Description\\ 
\hline
1 & $\neq 0$ & $<0$     & $\lambda_1\cdot\lambda_2<0$         & 3D X\\
2 & $>0$     & $>0$     & $\lambda_1=\lambda_2^*$             & 3D O (Repelling)\\
3 & $<0$     & $>0$     & $\lambda_1=\lambda_2^*$             & 3D O (Attracting)\\
4 & $\neq 0$ & $\leq 0$ & $\lambda_1,\lambda_2>0$             & 3D Repelling\\
5 & $\neq 0$ & $\leq 0$ & $\lambda_1,\lambda_2<0$             & 3D Attracting\\
6 & $\neq 0$ & $<0$     & $\lambda_1\ or\ \lambda_2=0$        & 3D Anti-parallel\\
7 & $=0$     & $<0$     & $\lambda_1=-\lambda_2\in\mathbb{R}$ & 2D X\\
8 & $=0$     & $>0$     & $\lambda_1=-\lambda_2\in\mathbb{C}$ & 2D O\\
9 & $=0$     & $=0$     & $\lambda_1=\lambda_2=0$             & 2D Anti-parallel\\
\hline
\end{tabular}
\end{table}

To investigated the geometric properties of $\mathbf{B}_\perp$, we further decompose $\mathbf{M}$ into two parts as
\begin{equation}
   \bf M=M'+T\,,\label{eq:dcompM}
\end{equation}
where,
\begin{align}
   \mathbf{M}' & = \left(\begin{array}{cc}
      p & \left(q-J_3\right)/2\\
      \left(q+J_3\right)/2    & -p 
   \end{array}\right)\,,\\
   \mathbf{T} & = \left(\begin{array}{cc}
      tr\left(\bf M\right)/2 & 0\\
      0    & tr\left(\bf M\right)/2
   \end{array}\right)\,,\\
   p & = \left(M_{11}-M_{22}\right)/2\,.
\end{align}
$\bf M'$ is a traceless matrix corresponding to the source-free part of $\mathbf{B}_\perp$ which can be treated as a real magnetic field, while $\bf T$ is the source part originating from the 3D effects. 
Correspondingly, $\mathbf{B}_\perp$ can be written as $\mathbf{B}_\perp = \mathbf{B}_\perp'+\nabla_pG$, where 
\begin{align}
   \mathbf{B}_\perp' & = \mathcal{J}\nabla_pF\,,\\
   F & = \frac{1}{4}\left[\left(q-J_3\right)R_2^2-\left(q+J_3\right)R_1^2\right]+pR_1R_2\,,\\
   G & = \frac{1}{4}tr\left(\bf M\right)\left(R_1^2+R_2^2\right)\,,\\
   \nabla_p & = \left(\partial/\partial R_1,\partial/\partial R_2\right)^T\,,
\end{align}
and
\begin{equation}
   \mathcal{J} = \left(\begin{array}{cc}
      0 & 1\\
      -1 & 0 
   \end{array}\right)\,\\
\end{equation}
is the symplectic matrix.
Because the second term $\nabla_pG$ is isotropic, the anisotropic properties of $\mathbf{B}_\perp$ are completely determined by the first term $\mathbf{B}_\perp'$.
As a 2D magnetic field, the integration curves of $\mathbf{B}_\perp'$ are contour lines of its Hamiltonian $F$.

\begin{figure}
\centering
\resizebox{\hsize}{!}{\includegraphics{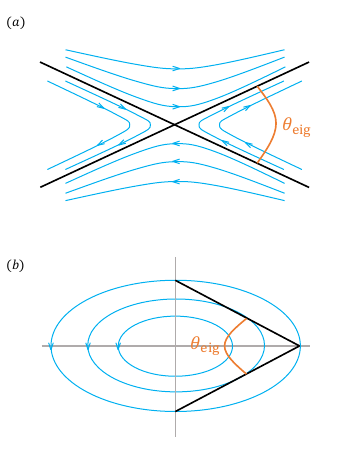}}
\caption{Cartoons of $\mathbf{B}_\perp'$ lines of X (a) and O (b) types. 
The blue curves depict field lines.
The definitions of $\theta_\mathrm{eig}$ are labeled by orange markers.}
\label{fig:eigangle}
\end{figure}

Following the method by \cite{Parnell1996}, we define a new coordinate $\mathbf{X}=\left(X_1,X_2\right)^T$ satisfying
\begin{equation}
   \mathbf{X} = \left(\begin{array}{cc}
      \cos\theta & -\sin\theta\\
      \sin\theta & \cos\theta\\
   \end{array}\right)\mathbf{R}\,,
\end{equation}
where the rotation angle $\theta$ satisfies $\tan2\theta=-2p/q$.
Then $F$ can be transformed to
\begin{equation}
   F = \frac{1}{4}\left[X_2^2\left(J_\mathrm{thres}-J_3\right)-X_1^2\left(J_\mathrm{thres}+J_3\right)\right]\,.\\
\end{equation}
Therefore, the field lines of $\mathbf{B}'$ are hyperbolic lines if $\left|J_3\right|<J_\mathrm{thres}$ (Fig.\,\ref{fig:eigangle}a), are concentric elliptical curves if $\left|J_3\right|>J_\mathrm{thres}$ (Fig.\,\ref{fig:eigangle}b), and are parallel straight lines if $\left|J_3\right|=J_\mathrm{thres}$.

To quantitatively evaluate the geometric shapes of $\mathbf{B}_\perp'$ lines, we define an angle,
\begin{align}
  \theta_\mathrm{eig} & = \begin{cases}
\arctan\left(\sqrt{\left(J_\mathrm{thres}/J_3\right)^2-1}\right)\,, & \left|J_3\right|<J_\mathrm{thres}\,,\\
\arctan\left(\sqrt{\left(J_3/J_\mathrm{thres}\right)^2-1}\right)\,, & \left|J_3\right|>J_\mathrm{thres}\,.
  \end{cases} 
\end{align}
For $\left|J_3\right|<J_\mathrm{thres}$, $\theta_\mathrm{eig}$ is the acute angle spanned by the separatrix lines (see Fig.\,\ref{fig:eigangle}a).
For $\left|J_3\right|>J_\mathrm{thres}$, $\theta_\mathrm{eig}$ is the angle spanned by two minor-axis vertices and one major-axis vertex of an elliptical curve (see Fig.\,\ref{fig:eigangle}b).
According to the definition, $\theta_\mathrm{eig}\in\left[0,90^\circ\right]$.
If $J_3\to J_\mathrm{thres}$, then $\theta_\mathrm{eig}\to 0$, which corresponds to the anti-parallel case.
Meanwhile, $\theta_\mathrm{eig}\to 90^\circ$ means that the X-type $\mathbf{B}'_\perp$ tends to be a potential field satisfying $J_3\to 0$ while the O-type $\mathbf{B}'_\perp$ represents as concentric circle lines governed only by the anti-symmetric part of $\mathbf{M}'$ (the parallel current $J_3$); 

Because $\theta_\mathrm{eig}$ only reflects the property of the source-free part $\bf M'$, we define another parameter to evaluate the importance of $\bf T$, namely the trace ratio 
\begin{equation}
R_\mathrm{tr}=\| \mathbf{T}\| /\| \mathbf{M}'\|\,,
\end{equation}
where $\|\cdot\|$ is the Euclidean norm.
Larger $R_\mathrm{tr}$ means larger effects from the trace part.

\section{Algorithm \label{sec:method}}

Combining the theory of general reconnection and the analysis of local magnetic structures, we develop a numerical method to locally analyze 3D reconnection sites.
When introducing this algorithm, we suppose the discrete data are uniformly sampled in a Cartesian frame, which, however, can be generalized into arbitrary coordinate systems via transformations.

\subsection{Step 1: Extract Grids with Large $\left|E_\parallel\right|$\label{ssec:step1}}

As discussed in Sect.\,\ref{ssec:GMRLocal}, we can select grids with finite values of $E_\parallel$ to represent the general reconnection sites.
But, for numerical data, $E_\parallel$ can hardly be zero.
Therefore, one can define a threshold value of $E_\mathrm{thres}>0$ to extract a subset of $\left|E_\parallel\right|>E_\mathrm{thres}$ from the entire domain to reduce the computational costs of the following procedures.
However, if the data scale is acceptable or $E_\mathrm{thres}$ is difficult to choose, to guarantee statistical completeness, one can also skip this step and analyze all grids.

\subsection{Step 2: Analyze Local Magnetic Structures}

First, approximate the 3D magnetic Jacobian matrix $\mathbf{D}_B$ on each grid via two-order central difference as
\begin{equation}
   \mathbf{D}_B^{lmn}=\left(\frac{\mathbf{B}^{l+1,m,n}-\mathbf{B}^{l-1,m,n}}{2\Delta x^1},\frac{\mathbf{B}^{l,m+1,n}-\mathbf{B}^{l,m-1,n}}{2\Delta x^2},\frac{\mathbf{B}^{l,m,n+1}-\mathbf{B}^{l,m,n-1}}{2\Delta x^3}\right)\,,
\end{equation} 
where $l$, $m$, and $n$ are indices of $x^1$, $x^2$, and $x^3$, respectively.
$\mathbf{B}^{lmn}=\left(B^{1,lmn},B^{2,lmn},B^{3,lmn}\right)^T$ is the magnetic field at $\mathbf{r}^{lmn}$.
Generally speaking, the two-order spatial difference is accurate enough for most cases, but one can replace it with higher-order schemes if necessary.

\begin{figure}
\centering
\resizebox{\hsize}{!}{\includegraphics{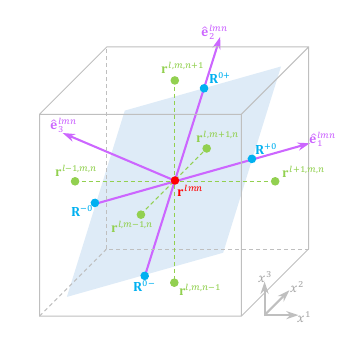}}
\caption{Schematic diagram of the local frame at a grid $\mathbf{r}^{lmn}$.
The purple arrow lines span the local frame.
The light blue shade presents the MPP.
The blue dots depict four adjacent grids on the MPP that are one step-length ($\Delta L$) away from the origin point.
The green dots are six adjacent grids surrounding $\mathbf{r}^{lmn}$ in the frame of original discrete data.
}
\label{fig:localframe}
\end{figure}

Second, construct local frame at $\mathbf{r}^{lmn}$ (see Fig.\,\ref{fig:localframe}).
The base vector on the third direction (the direction of the local magnetic field) is $\hat{\bf e}_3^{lmn}=\left(B^{1,lmn},B^{2,lmn},B^{3,lmn}\right)/\left|\mathbf{B}^{lmn}\right|$.
$\hat{\bf e}_2^{lmn}$ can be any unit vector perpendicular to $\hat{\bf e}_3^{lmn}$.
As a convenient choice, we set it as $\hat{\bf e}_2^{lmn}=\left(B^{2,lmn},-B^{1,lmn},0\right)/\sqrt{\left(B^{1,lmn}\right)^2+\left(B^{2,lmn}\right)^2}$. 
For the special case of $B^{1,lmn}=B^{2,lmn}=0$, let $\hat{\bf e}_2^{lmn}=\left(0,1,0\right)$.
The last base vector can be directly obtained by $\hat{\bf e}_1^{lmn}=\hat{\bf e}_2^{lmn}\times\hat{\bf e}_3^{lmn}$.
Hereafter, the local plane spanned by $\hat{\bf e}_1^{lmn}$ and $\hat{\bf e}_2^{lmn}$ within a cell is called the magnetic projection plane (MPP).

Third, transform $\mathbf{D}_B^{lmn}$ into the local frame. 
Because both the original and the local frames are Cartesian frames and the base vector systems are orthogonal normalized ones, the transformation matrix can be proven to be
\begin{equation}
   \mathcal{T}^{lmn} = \left(\begin{array}{c}
      \hat{\bf e}_{1}^{lmn}\\
      \hat{\bf e}_{2}^{lmn}\\
      \hat{\bf e}_{3}^{lmn}
   \end{array}\right)\,,\label{eq:Tlocal}
\end{equation}
and its inverse matrix is $\mathcal{T}^{lmn}_{\mathrm{inv}}=\left(\mathcal{T}^{lmn}\right)^T$.
In the local frame, $\mathbf{D}_B^{lmn}$ is transformed to $\mathbf{D'}_B^{lmn}=\mathcal{T}^{lmn}\mathbf{D}^{lmn}\mathcal{T}^{lmn}_\mathrm{inv}$.

Four, analyze the local magnetic structures via calculating the trace, discriminant, eigenvalues, $\theta_\mathrm{eig}$, and $R_\mathrm{tr}$ of the matrix
\begin{equation}
   \mathbf{M}^{lmn} = \left(\begin{array}{cc}
      D_{B,11}'^{lmn} & D_{B,12}'^{lmn}\\
      D_{B,21}'^{lmn} & D_{B,22}'^{lmn}
   \end{array}\right)\,,\label{eq:M2DMD}
\end{equation}
as discussed in Sect.\,\ref{ssec:topology}.

\subsection{Step 3: Locate Reconnection Sites of extremal $E_\parallel$}

As discussed in Sect.\,\ref{ssec:GMRLocal}, we further locate the positions with extremal values of $E_\parallel$.
Theoretically, they can be identified by $\nabla_\perp E_\parallel=0$.
However, numerically, $\nabla_\perp E_\parallel$ can hardly vanish and it is also difficult to define a general small-value threshold for arbitrary cases.
Thus, to approximately locate these sites, we directly judge whether $\mathbf{r}^{lmn}$ is a 2D extremal point of $E_\parallel$ on the MPP. 
To implement this, four grids adjacent with $\mathbf{r}^{lmn}$ (see Fig.\,\ref{fig:localframe}) on the MPP is defined as,
\begin{align}
   \mathbf{R}^{-0} & = \mathbf{r}^{lmn}-\Delta L\hat{\bf e}_1^{lmn}\,,\\
   \mathbf{R}^{+0} & = \mathbf{r}^{lmn}+\Delta L\hat{\bf e}_1^{lmn}\,,\\
   \mathbf{R}^{0-} & = \mathbf{r}^{lmn}-\Delta L\hat{\bf e}_2^{lmn}\,,\\
   \mathbf{R}^{0+} & = \mathbf{r}^{lmn}+\Delta L\hat{\bf e}_2^{lmn}\,,
\end{align}
where $\Delta L$ is the spatial step length of the original discrete data.
The values of $E_\parallel$ at these positions can be obtained via linear interpolation of original data.
If the values of $E_\parallel$ have the same signs at $\mathbf{r}$, $\mathbf{R}^{-0}$, $\mathbf{R}^{+0}$, $\mathbf{R}^{0-}$, and $\mathbf{R}^{0+}$, and $\left|E_\parallel\right|$ at $\mathbf{r}^{lmn}$ is maximum among the five positions, then we identify $\mathbf{r}^{lmn}$ as a 2D extreme point of $\left|E_\parallel\right|$.

\begin{table*}
\caption{Key configuration fields of the \texttt{Parameters} structure.}
\label{tab:lordconf}
\centering
\begin{tabular}{c c c}
\hline\hline 
Fields & Description & Values \\ 
\hline
\texttt{ARD\_AnalyzeAllGrids} & Skip Step 1 and analyze all grids & 0: false; 1: true\\
\texttt{ARD\_ScalarThreshold} & Threshold value of $\left|E_\parallel\right|$ & Nonnegative real numbers\\
\texttt{ARD\_ShowThresScalarProfile} & Draw a histogram of $\left|E_\parallel\right|$ without running \texttt{ARD}  & 0: false; 1: true\\
\texttt{ARD\_FixTrace} & Fix the trace of $\mathbf{D}^{lmn}_B$ from numerical errors & 0: false; 1: true\\
\texttt{ARD\_AnalyzeLocalEffects} & Analyze local reconnection effects (see Appendix \ref{asec:localeffects}) & 0: false; 1: true\\
\texttt{NumRAMBlock} & Number of RAM blocks & Positive integer \\
\texttt{OutputType} & Select format of output file & -1: no output file; 0: .mat; 1: .csv \\
\texttt{OutputDir} & Output directory & String  \\
\texttt{OutputLabel} & User-defined label on output filename  & String \\
\texttt{OutputExtraData} & Output \texttt{ExtraData} & 0: false; 1: true  \\
\hline
\end{tabular}
\end{table*}

\section{The \texttt{ARD} function in \textsf{LoRD} Toolkit\label{sec:LoRD}}
This method has been implemented as an easy-to-use function \texttt{ARD}, integrated into our open-source \textsf{Matlab} toolkit project named \textsf{LoRD} (Locate Reconnection Distribution).
\textsf{LoRD} can be freely downloaded by the GitHub link \texttt{ ``https://github.com/RainthunderWYL/LoRD.git''}.
To use the functions in \textsf{LoRD}, one just needs to add the directory into the \textsf{Matlab} environment by the following command:
\begin{lstlisting}[language={matlab}]
   path('<Download Directory>/LoRD/matlab',path);
\end{lstlisting}

\subsection{The \texttt{ARD} function} 

The \texttt{ARD} function can be executed as follows: 
\begin{lstlisting}[language={matlab}]
   RDInfo = ARD(B1,B2,B3,x1,x2,x3,Parameters,N1,N2,N3);
\end{lstlisting}
Here, \texttt{B1}, \texttt{B2}, and \texttt{B3} are 3D matrices of discrete magnetic field on three directions.
Notice that they follow the ``meshgrid'' data model of \textsf{Matlab}, namely, the first dimension is $x^2$, the second dimension is $x^1$, and the third dimension is $x^3$.
\texttt{x1}, \texttt{x2}, and \texttt{x3} are 1D coordinate arrays on three directions.
At present, \texttt{ARD} can only process uniform Cartesian mesh.
But users can use interpolation to generate uniform Cartesian mesh inputs from other complex mesh models.
\texttt{N1}, \texttt{N2}, and \texttt{N3} are three optional inputs with the same size as the magnetic field data, which passes user-defined data of $\bf N$ to \texttt{ARD}.
Without \texttt{N1}--\texttt{N3}, \texttt{ARD} will suppose the nonideal term is induced by a Joule dissipation with a constant resistivity and use $\mathbf{J}=\nabla\times\mathbf{B}$ to replace $\bf N$.

\texttt{Parameters} is a structure containing fields of key configuration parameters as listed in Table\,\ref{tab:lordconf}.
If \texttt{Parameters.ARD\_AnalyzeAllGrids} is set as 1, \texttt{ARD} skips Step 1 (see Sect.\,\ref{ssec:step1}) and analyzes all girds, which should be carefully considered because of the considerable costs of computational resources.
The threshold value of $\left|E_\parallel\right|$ is defined by \texttt{Parameters.ARD\_ScalarThreshold}.
To help users determine this value, \texttt{ARD} can present a histogram of $\left|E_\parallel\right|$ without proceeding further analysis if \texttt{Parameters.ARD\_ShowThresScalarProfile} is set as 1.
Because the magnetic Jacobian matrix $\mathbf{D}^{lmn}_B$ might have small trace components resulting from numerical errors, \texttt{ARD} provides an option to clean the trace by $\mathbf{D}^{lmn}_B-1/3tr\left(\mathbf{D}^{lmn}_B\right)\mathbf{I}$, which is controlled by \texttt{Parameters.ARD\_FixTrace}.
If \texttt{Parameters.NumRAMBlock} is larger than 1, \texttt{ARD} divides the original field data into \texttt{NumRAMBlock} blocks and sequentially analyzes each block to avoid RAM overflow, which is useful for processing massive magnetic field data.

The output \texttt{RDInfo} is a structure containing two fields, namely, \texttt{RDInfo.Data} and \texttt{RDInfo.ExtraData}.
\texttt{RDInfo.Data} is a $N_\mathrm{sl}\times 7$ matrix.
Each row saves the data of an analyzed grid and the columns correspond to 7 key attributes, including \texttt{x1}, \texttt{x2}, \texttt{x3}, \texttt{RDType}, \texttt{Is2DExtrema}, \texttt{EigAngle}, and \texttt{RatioMTrace}.
\texttt{x1}, \texttt{x2}, and \texttt{x3} are the $x^1$, $x^2$, and $x^3$ coordinates, respectively.
\texttt{RDType} is an integer number taking values from 1 to 9, marking the types of $\mathbf{B}_\perp$ as listed in Table\,\ref{tab:types}.
\texttt{Is2DExtrema} labels whether the grid is a 2D extreme point of $E_\parallel$.
\texttt{EigAngle} and \texttt{RatioMTrace} save the values of $\theta_\mathrm{eig}$ and $R_\mathrm{tr}$, respectively.
\texttt{RDInfo.ExtraData} is an optional output controlled by \texttt{Parameters.OutputExtraData}, which has 25 columns containing information of $\left|\mathbf{B}^{lmn}\right|$, $\mathbf{D}'^{lmn}_B$, $\hat{\bf e}_1^{lmn}$, $\hat{\bf e}_2^{lmn}$, $\hat{\bf e}_3^{lmn}$, and parameters for evaluating local reconnection effects (see Table \ref{tab:lordextradata}).
It should be noticed that \texttt{Parameters.OutputExtraData} is forcibly set as 1 if \texttt{Parameters.ARD\_AnalyzeLocalEffects} is enabled.
\texttt{ARD} can also save the output \texttt{RDInfo} into ``.mat'' or ``.csv'' files, which is set by \texttt{Parameters.OutputType}.
Users can customize the output directory by \texttt{Parameters.OutputDir} and also attach a label behind the filename via \texttt{Parameters.OutputLabel}.

\begin{table}
\caption{Data saved in \texttt{RDInfo.ExtraData}. $\bf \Gamma$ is defined in Appendix \ref{asec:localeffects}.}
\label{tab:lordextradata}
\centering
\begin{tabular}{c c c}
\hline\hline 
Column No. & Index Name & Physical Symbol \\ 
\hline
1 & \texttt{B0} & $\left|\mathbf{B}\right|$\\
2 & \texttt{DB11} & $D'_{B,11}$\\
3 & \texttt{DB12} & $D'_{B,12}$\\
4 & \texttt{DB21} & $D'_{B,21}$\\
5 & \texttt{DB22} & $D'_{B,22}$\\
6 & \texttt{DB31} & $D'_{B,31}$\\
7 & \texttt{DB32} & $D'_{B,32}$\\
8 & \texttt{DB13} & $D'_{B,13}$\\
9 & \texttt{DB23} & $D'_{B,23}$\\
10 & \texttt{DB33} & $D'_{B,33}$\\
11 & \texttt{e11} & $\hat{\mathrm{e}}_{1,1}$\\
12 & \texttt{e12} & $\hat{\mathrm{e}}_{1,2}$\\
13 & \texttt{e13} & $\hat{\mathrm{e}}_{1,3}$\\
14 & \texttt{e21} & $\hat{\mathrm{e}}_{2,1}$\\
15 & \texttt{e22} & $\hat{\mathrm{e}}_{2,2}$\\
16 & \texttt{e23} & $\hat{\mathrm{e}}_{2,3}$\\
17 & \texttt{e31} & $\hat{\mathrm{e}}_{3,1}$\\
18 & \texttt{e32} & $\hat{\mathrm{e}}_{3,2}$\\
19 & \texttt{e33} & $\hat{\mathrm{e}}_{3,3}$\\
20 & \texttt{DNpara1} & $\partial E_\parallel/\partial R_1$\\
21 & \texttt{DNpara2} & $\partial E_\parallel/\partial R_2$\\
22 & \texttt{Gamma1} & $\Gamma_1$\\
23 & \texttt{Gamma2} & $\Gamma_2$\\
24 & \texttt{CurlN} & $\left|\nabla\times\mathbf{N}\right|$\\
25 & \texttt{BxCurlN} & $\left|\mathbf{B}\times\left(\nabla\times\mathbf{N}\right)\right|$\\
\hline
\end{tabular}
\end{table}

\subsection{Input API}

We implement an input API function for loading the binary data files of discrete fields to simplify the usage of \texttt{ARD} for users unfamiliar with \textsf{Matlab}.
For example, to load an $n_1\times n_2\times n_3$ data $B^1$ from a binary file with ``double'' precision named ``B1.bin'', one can call the following commands:
\begin{lstlisting}[language={matlab}]
   DIM = [n1,n2,n3];
   Precision = 'double';
   LIM = [x1s,x1e,x2s,x2e,x3s,x3e];
   [B1,x1,x2,x3] = Tool_LoadData_Bin('B1.bin',DIM,Precision,LIM);
\end{lstlisting}
The binary file should save the 3D data of $B^1$ in a 1D array, in which the $x^1$-direction index changes fastest and the $x^3$-direction index changes slowest.
\texttt{Precision} is a string setting the class and size of bits to read which should be consistent with the style of the binary file.
Available parameters of \texttt{Precision} include \texttt{`double'}, \texttt{`float32'}, etc..
\texttt{LIM} is an optional parameter defining the coordinate limitations in three directions.
Here, \texttt{x1s} and \texttt{x1e} define the start and end of $x^1$-coordinate, respectively, and \texttt{x2s}, \texttt{x2e}, \texttt{x3s}, and \texttt{x3e} have similar definitions.
Without \texttt{LIM}, this function will simply assign the outputs \texttt{x1}, \texttt{x2}, and \texttt{x3} with integer grid indices.
The output variables \texttt{B1}, \texttt{x1}, \texttt{x2}, and \texttt{x3} can be directly passed to the \texttt{ARD} function.

\subsection{Standards of Code Upgrade}

The functions and configuration parameters of the \texttt{ARD} code introduced above are based on the current version.
With the development of our method, the code might also be upgraded correspondingly.
For example, more output parameters might be defined to give a more comprehensive picture of reconnection sites and the Matlab scripts will also be optimized for better efficiency.
However, the code upgrade will follow a backward-compatible standard.
To be specific, the I/O APIs will remain unchanged.
If necessary, we will add new configuration parameters into the \texttt{Parameters} structure or output new reconnection variables in the \texttt{RDInfo} structure.
All new features of our code will be updated in time on the GitHub website in the future.

\begin{figure*}
\centering
\includegraphics[width=1\textwidth]{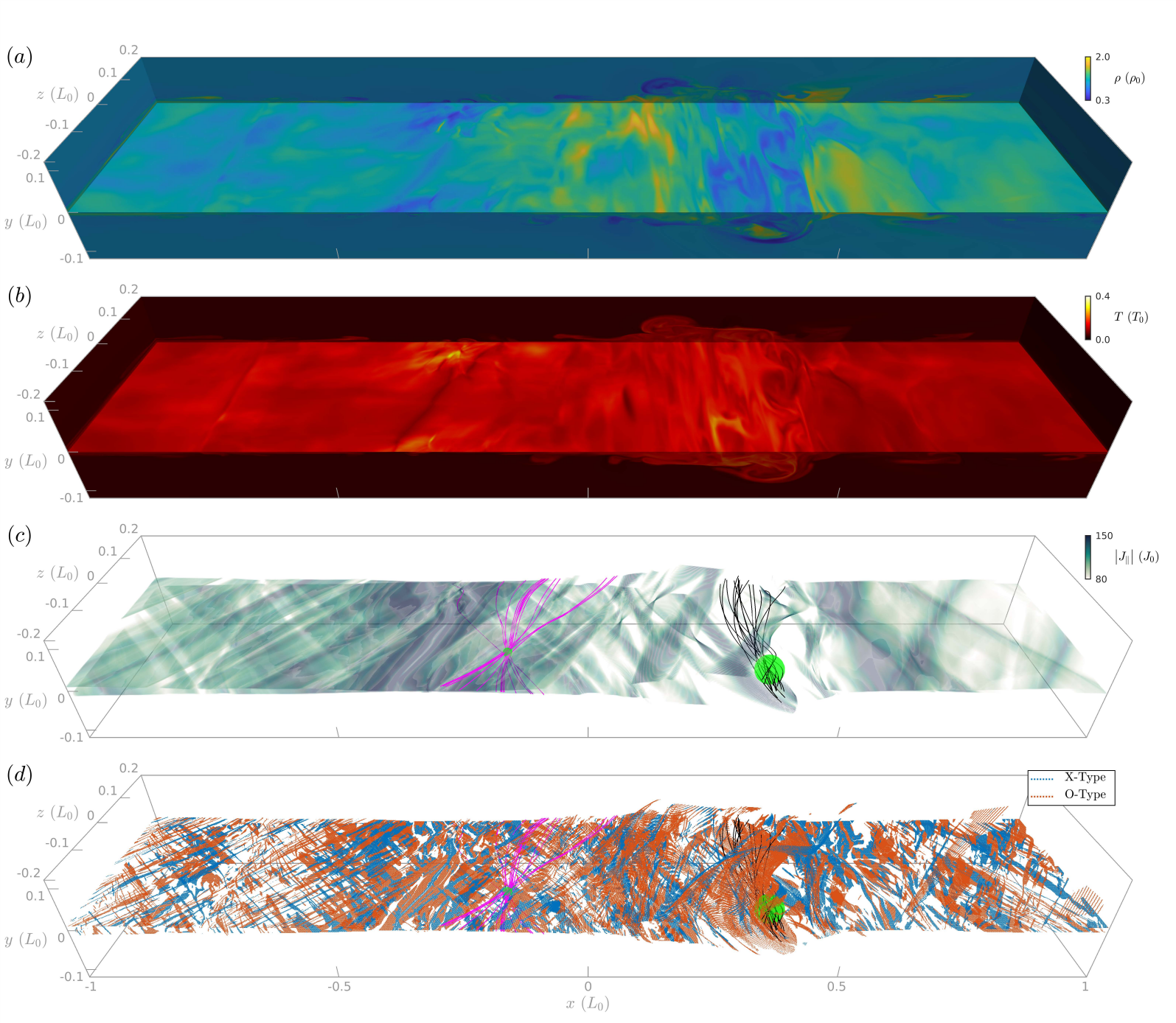}
\caption{The distributions of mass density $\rho$ (a), temperature $T$ (b), parallel current density $\left|J_\parallel\right|$ (c), and 3D X/O-type grids (d) at $t=9$.
For $\rho$ and $T$, we only depict their distributions on the current-sheet middle plane ($y=0$), the front/back planes ($z=\pm 0.2$), and the left/right planes ($x=\pm 1$).
Panel (c) shows the 3D profile of $\left|J_\parallel\right|$ satisfying $\left|J_\parallel\right|>80$.
In panel (d), we plot the grids that (1) satisfy $\left|J_\parallel\right|>50$, (2) have local magnetic structures of 3D X and O types, and (3) are 2D projected extreme points of $J_\parallel$.
The X-type (type 1) grids are depicted by blue dots, while the O-type grids (types 2 and 3) are plotted by orange ones. 
The magenta and black curves in panels (c) and (d) are examples of sheared and twisted field lines, which are traced from the initial sampling positions near 3D X-type and O-type grids, respectively.
The green spheres mark the regions for sampling start positions of field-line tracing.
An animation of this figure showing the entire evolution from $t=0$ to $9$ is available.
}
\label{fig:hs3d}
\end{figure*}

\begin{figure*}
\centering
\includegraphics[width=1\textwidth]{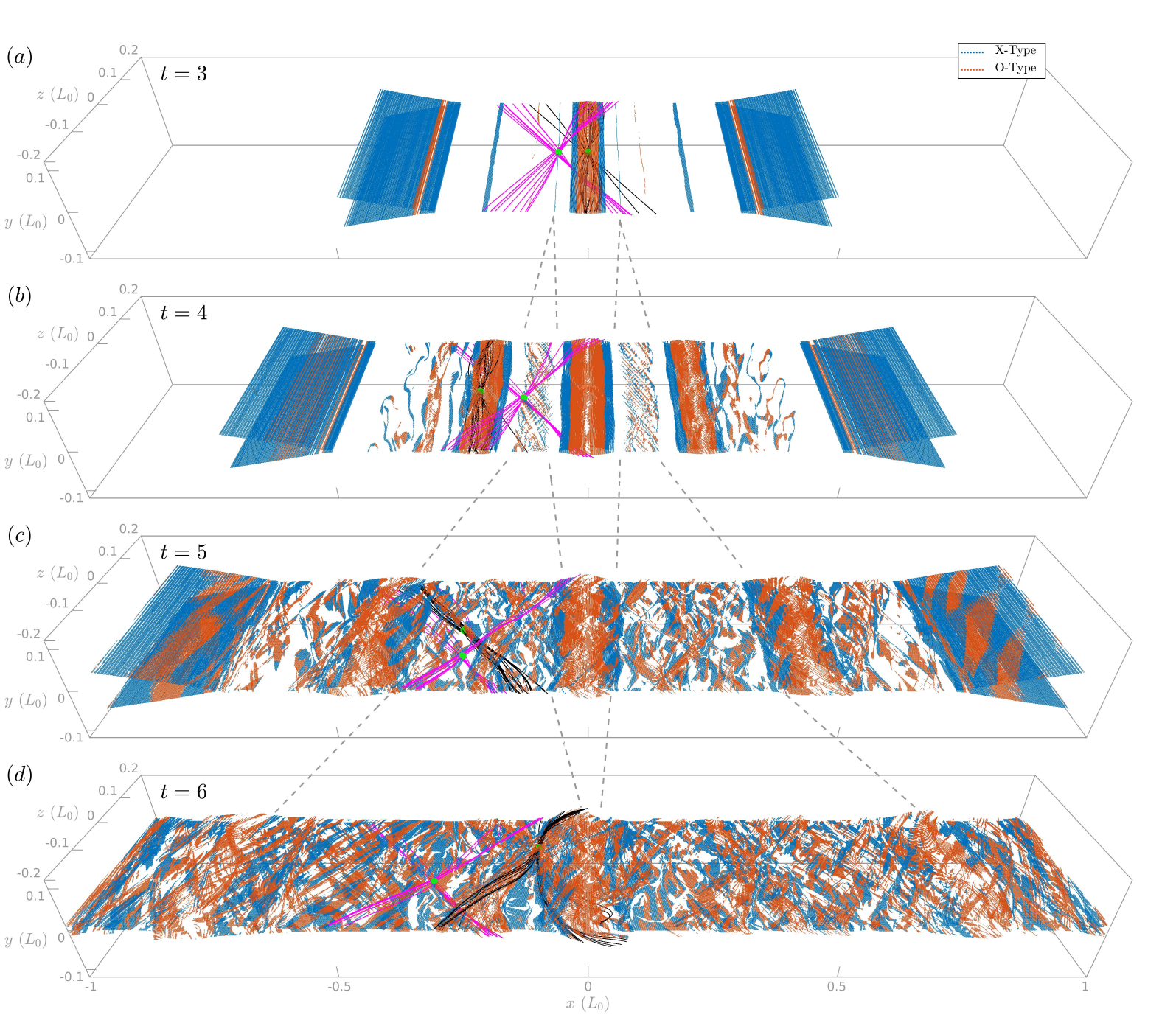}
\caption{Four snapshots of 3D X/O-type grids exhibiting the development of turbulent reconnection.
The gray dashed lines approximately trace the origin of turbulent reconnection regions.
}
\label{fig:hs3d_ev}
\end{figure*}

\begin{figure}
\centering
\resizebox{\hsize}{!}{\includegraphics{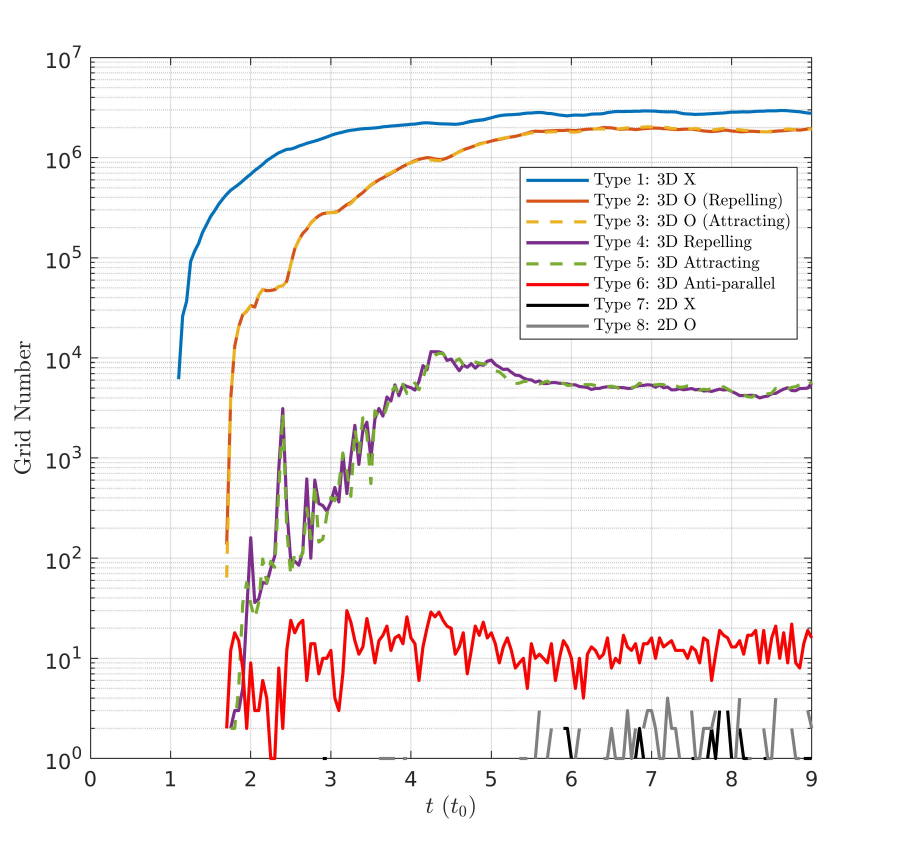}}
\caption{The evolutions of grid numbers for different types of structures.
Only grids with $\left|J_\parallel\right|>50$ are counted.
}
\label{fig:hist_types}
\end{figure}

\begin{figure*}
\centering
\includegraphics[width=0.6\textwidth]{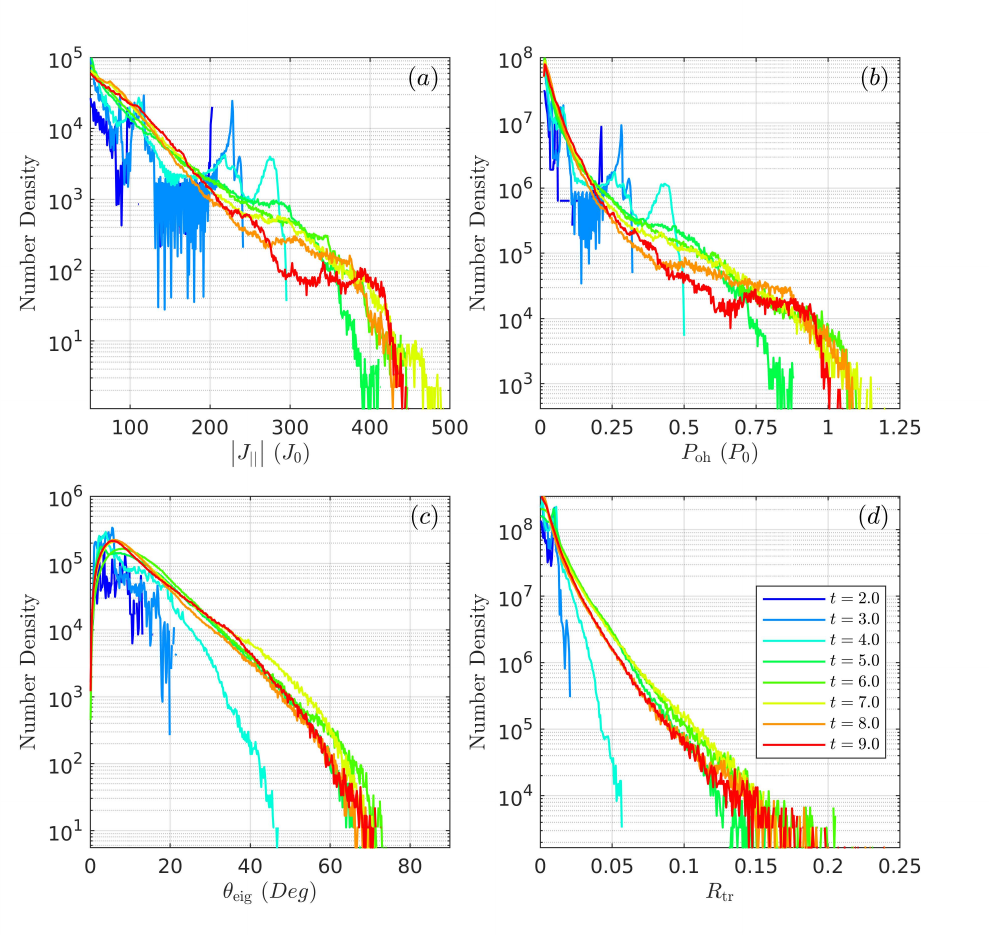}
\caption{The number densities of $\left|J_\parallel\right|$ (a) , $P_\mathrm{oh}$ (b), $\theta_\mathrm{eig}$ (c), and $R_\mathrm{tr}$ (d) of the 3D X-type grids at different moments.
Different colors correspond to different moments.
}
\label{fig:hist_x_alltime}
\end{figure*}

\begin{figure*}
\centering
\includegraphics[width=0.6\textwidth]{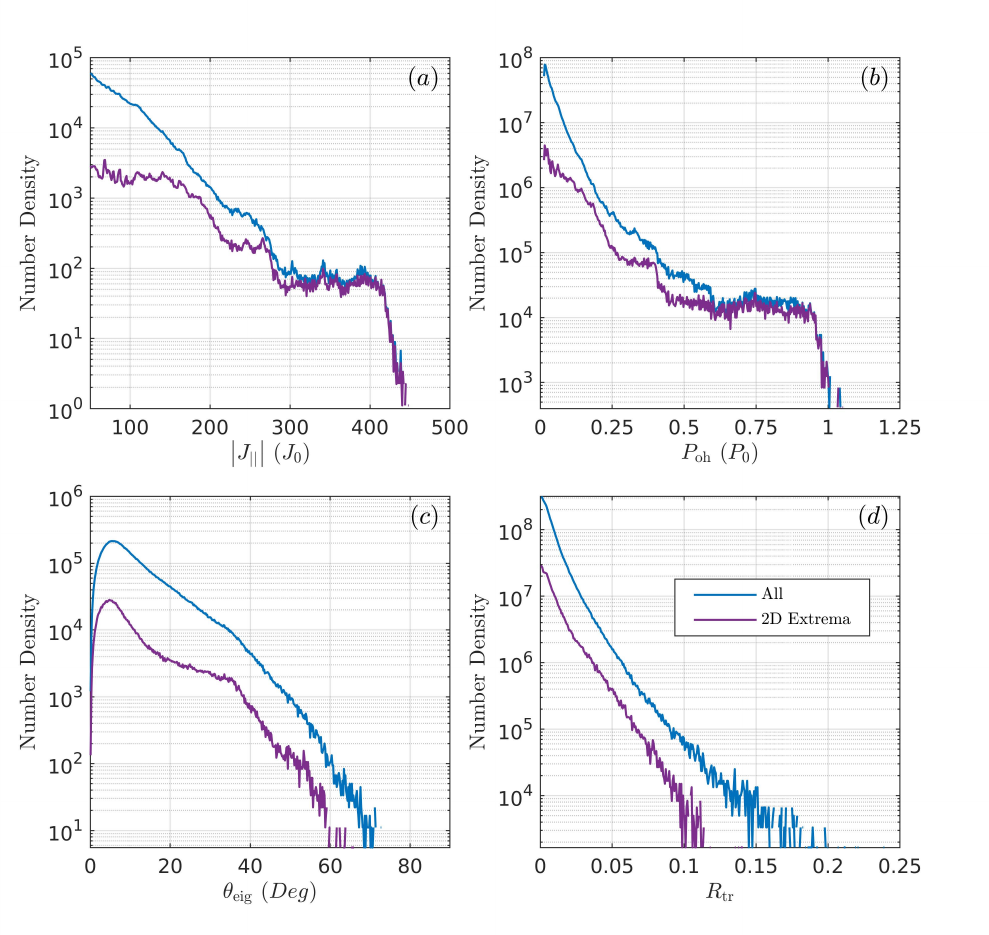}
\caption{The number densities of $\left|J_\parallel\right|$ (a), $P_\mathrm{oh}$ (b), $\theta_\mathrm{eig}$ (c), and $R_\mathrm{tr}$ (d) of the 3D X-type grids at $t=9$.
The blue curves depict profiles of all 3D X-types grids, while the purple ones plot the subsets with 2D extrema of $J_\parallel$.
}
\label{fig:hist_x_180}
\end{figure*}

\section{Benchmark 1: 3D Turbulent Reconnection within a Harris Sheet\label{sec:test1}}

\subsection{Numerical Model}
To test the validation of our method, we perform a 3D MHD simulation of magnetic reconnection within a Harris sheet.
We solve the resistive MHD equations:
\begin{align}
\frac{\partial\rho}{\partial t}+\nabla\cdot\left(\rho{\bf u}\right) & = 0\,,\nonumber \\
\frac{\partial\left(\rho{\bf u}\right)}{\partial t}+\nabla\cdot\left(\rho{\bf u}{\bf u}-{\bf BB}+P^{*}{\bf I}\right) & = 0\,,\nonumber \\
\frac{\partial e}{\partial t}+\nabla\cdot\left[\left(e+P^{*}\right){\bf u}-{\bf B}\left(\bf{B}\cdot{\bf u}\right)\right] & = 0\,, \label{eq:MHD}\\
\frac{\partial{\bf B}}{\partial t}-\nabla\times\left(\bf{u}\times\bf{B}\right) & = -\nabla\times\left(\eta{\bf J}\right)\,,\nonumber \\
{\bf J} & = \nabla\times{\bf B}\,,\nonumber
\end{align}
where, $\rho$, $\bf{u}$, $p$, $\bf{B}$, $T$, and $\bf{J}$ denote plasma density, velocity, thermal pressure, magnetic field, temperature, and current density, respectively.
$P^*$ equals $p+B^2/2$ and $\bf{I}$ is the identity matrix.
$e=p/\left(\gamma-1\right)+\rho u^2/2+B^2/2$ is the total energy, where $\gamma=5/3$ is the adiabatic index.
All physical quantities are normalized based on the dimensionless units as listed in Table\,\ref{tab:units}.

\begin{table}
\caption{Dimensionless Units}
\label{tab:units}
\centering
\begin{tabular}{c c c}
\hline\hline 
Unit of& Symbol & Value \\ 
\hline
Mass                  & $\bar{m}$        & $8.36\times 10^{-25}\,\mathrm{g}$\\
Number density        & $n_0$            & $2\times 10^{10}\,\mathrm{cm^{-3}}$\\
Space                 & $L_0$            & $5\times 10^{9}\,\mathrm{cm}$\\
Magnetic strength     & $B_0$            & $20\,\mathrm{G}$\\
Mass density          & $\rho_0$         & $1.67\times 10^{-14}\,\mathrm{g\,cm^{-3}}$\\
Time                  & $t_0$            & $114.61\,\mathrm{s}$\\
Velocity              & $u_0$            & $4.36\times 10^{7}\,\mathrm{cm\,s^{-1}}$\\
Temperature           & $T_0$            & $1.15\times 10^{7}\,\mathrm{K}$\\
Pressure              & $p_0$            & $31.8\,\mathrm{dyn\,cm^{-2}}$\\
Energy density        & $e_0$            & $31.8\,\mathrm{erg\,cm^{-3}}$\\
Current density       & $J_0$            & $9.54\,\mathrm{statC\,s^{-1}\,cm^{-2}}$\\
\hline
\end{tabular}
\end{table}

The initial magnetic field forms a force-free Harris sheet, namely,
\begin{align}
B_{x} & = B_0\tanh\left(y/\lambda\right)\,,\nonumber \\
B_{y} & = 0\,,\label{eq:mag}\\
B_{z} & = B_0/\cosh\left(y/\lambda\right)\,,\nonumber
\end{align}
where $B_0=1$ and $\lambda=0.1$ is the half-width of the current sheet.
The initial mass density and pressure are uniformly set as $1$ and $0.05$, respectively, corresponding to a $\beta=0.1$.
We set a uniform background resistivity $\eta=5\times10^{-6}$ to obtain a Lundquist number $S=L_0u_0/\eta=2\times 10^5$.
To trigger fast reconnection, a perturbation on the $z$-direction of the magnetic vector field is placed at the center of the current sheet \citep[also see][]{Ye2020}, namely,
\begin{equation}
   \tilde{A}_z=A_p\exp\left(-\frac{x^2+y^2}{2w_A^2}\right)\,,
\end{equation}
where, $A_p=0.03$ and $w_A=0.1$.
The initial velocity field is a small-amplitude Gaussian thermal noise with a zero mean value and a standard deviation $\sigma_u=10^{-3}$, which works as a seed of the self-sustained turbulence \citep[also see][]{Huang2016}.

The simulation domain is $x\in\left[-2.5,2.5\right]$, $y\in\left[-0.5,0.5\right]$, and $z\in\left[-0.2,0.2\right]$.
The $z$-boundaries are periodic, while the rest are open boundaries.
The static mesh refinement technique is applied to implement a uniform mesh in the reconnection region and save computational costs.
The root level-0 grid numbers are set as $1200$, $240$, and $96$ on $x$, $y$, and $z$ directions, respectively.
The level-1 grid number doubles in three directions in the region $0.1<\left|y\right|<0.3$.
The level-2 grid, doubling again, is set in the core reconnection region $\left|y\right|\leq 0.1$, implementing an effective mesh of $4800\times 960\times 384$ there.

We solve the above system using the \textsf{Athena++} code \citep{Stone2020}.
The HLLD Riemann solver \citep{Miyoshi2005}, the 2-order piecewise linear method (PLM), and the 2-order van Leer predictor-corrector scheme are selected for solving the conservative part of the MHD equations.
The resistivity term is calculated by the explicit operator splitting method.
The 2-order RKL2 super-time-stepping algorithm is applied to reduce computational costs \citep{Meyer2014}.
The simulation stops at $t=9$.

\subsection{Evolution Overview}

The evolution presents a standard picture of the Harris-sheet reconnection (see the animation of Fig.\,\ref{fig:hs3d}).
In the beginning, triggered by the perturbation field $\tilde{A}_z$, the center of the current sheet first gets thinner, where the density, temperature, and current also increase simultaneously.
During this stage, the system shows a translate symmetry on $z$-direction and tends to the 2D results.
After $t=3$, 3D tearing-mode instability starts to dominate the evolution and the reconnection is significantly boosted.
Various small-scale structures carrying strong currents appear and the distributions of density and temperature get highly nonuniform in three directions.
After $t=6$,  the entire current sheet reaches a fully-developed turbulent pattern wherein the oblique tearing-mode fluctuations can be recognized \citep[also see][]{Huang2016}.
Finally, at $t=9$, several relatively large-scale vortex-like structures containing complex density and temperature profiles grow near the center of the current sheet (see Fig.\,\ref{fig:hs3d}a and b).

\subsection{Qualitative Global Picture}

As discussed in Sect.\,\ref{ssec:GMR}, the regions with strong $E_\parallel$ can reflect the locations of general reconnection, which can be equivalently replaced by $J_\parallel$ since we apply a uniform resistivity.
After the reconnection enters the turbulent stage, the regions with strong $J_\parallel$ show patchy-like patterns (see Fig.\,\ref{fig:hs3d}c).

Based on the algorithm in Sect.\,\ref{sec:method}, we obtain the information of magnetic structures at strong $J_\parallel$ regions.
In Fig.\,\ref{fig:hs3d}, to give a clear picture, we only plot 3D X/O-type grids having 2D projected extrema of $J_\parallel$ (see Fig.\,\ref{fig:hs3d}d).
The X-type grids are the locations undergoing the SL reconnection \citep[][]{Priest1989}.
Meanwhile, the O-type grids can approximately present positions of flux ropes consisting of twisted field lines.

The distributions of X/O-type grids can exhibit the dynamical evolution of reconnection regions (see Fig.\,\ref{fig:hs3d_ev}).
At $t=3$, several X-lines and O-lines start to emerge near the center region (Fig.\,\ref{fig:hs3d_ev}a).
Later, the X-lines are shattered by the tearing-mode instability and both X and O-type structures with oblique patterns are generated simultaneously (Fig.\,\ref{fig:hs3d_ev}b). 
The newborn reconnection regions keep extending on the $x$-direction and finally develop complex turbulent states (Fig.\,\ref{fig:hs3d_ev}c, d).

The local magnetic structures on the MPP can also present global features of a bundle of field lines near a grid.
In Fig.\,\ref{fig:hs3d}c and d, we trace several field lines starting from a local spherical volume near an X-type grid (see the magenta curves), which present a global sheared pattern.
As another example, the black field lines, initially sampled within a volume containing O-type grids, show a typical pattern of flux ropes, which corresponds to the vortex structure also observed in the density and temperature profiles (Fig.\,\ref{fig:hs3d}).
Similar examples of field lines are also exhibited in Fig.\,\ref{fig:hs3d_ev}.

\subsection{Quantitative Statistical Analysis}

To study the generation probabilities of magnetic structures with different structures, we count the grid numbers of different types (see Fig.\,\ref{fig:hist_types}).
At the early stage, the magnitudes of $\left|J_\parallel\right|$ on all grids are less than the threshold value and thus no grid is recognized.
After $t=1$, type-1 grids (3D X-type) first appear.
Later, their number increases rapidly and finally reaches a platform after $t=6$.
The numbers of type-2 and type-3 grids are almost the same, showing a similar evolution trend as type-1 grids.
The growth of O-type grids corresponds to the formation of twisted flux ropes and thus represents the development of the tearing-mode instability.
The numbers of type-4 and type-5 are also approximately the same but are at least 2-order less than types 1--3.
The type-6 grids, with the 3D anti-parallel lines, are very rare because zero eigenvalues can hardly appear within numerical data.
However, many grids recognized as types 1--5 are close to anti-parallel lines, because their $\theta_\mathrm{eig}$ is close to zero.
This fact explains the reason why we use $\theta_\mathrm{eig}$ as a key parameter of $\bf B_\perp$.
Finally, although several grids of 2D types are identified, their number is small enough to ignore.

Now we focus on the statistical properties of type-1 grids, which represent the evolution rules of the SL reconnection (see Fig.\,\ref{fig:hist_x_alltime}).
During $t<3$, though the relative errors of number density profiles are relatively large because of the small number of recognized grids, some reconnection information can still be reflected. 
To be specific, both $\left|J_\parallel\right|$ and $P_\mathrm{oh}=\eta J^2$ are small corresponding to a moderate reconnection; $\theta_\mathrm{eig}$ mainly distributes below $10^\circ$ meaning the reconnection sites tend to have magnetic structures that are close to anti-parallel; $R_\mathrm{tr}$ is close to 0 implying the 3D effect is weak.
With the development of turbulence, the number density curves of $\left|J_\parallel\right|$, $P_\mathrm{oh}$, $\theta_\mathrm{eig}$, and $R_\mathrm{tr}$ keep extending toward larger values (see Fig.\,\ref{fig:hist_x_alltime}).
At $t=9$, the number density of $\left|J_\parallel\right|$ first decreases monotonically as $\left|J_\parallel\right|$ increases to 300, then forms a platform at $\left|J_\parallel\right|\in\left[300,400\right]$, and finally rapidly damps to zero at $\left|J_\parallel\right|\sim 440$ (Fig.\,\ref{fig:hist_x_alltime}a).
The distribution of $P_\mathrm{oh}$ presents a similar trend with $\left|J_\parallel\right|$ (Fig.\,\ref{fig:hist_x_alltime}b).
$\theta_\mathrm{eig}$ mainly distributes in the region $\theta_\mathrm{eig}<40^\circ$ and its peak number density appears near $5^\circ$ (Fig.\,\ref{fig:hist_x_alltime}c).
The maximum value of $\theta_\mathrm{eig}$ reaches $70^\circ$ and there are still considerable grids as $\theta_\mathrm{eig}\to 0$.
The number density of $R_\mathrm{tr}$ decreases monotonically in the entire domain.
After 3D effects get significant, more grids with larger $R_\mathrm{tr}$ emerge (Fig.\,\ref{fig:hist_x_alltime}d).
It should be noticed that these results might be model-dependent and should be reconsidered for different simulations.
For instance, the values and behaviors of $\left|J_\parallel\right|$ and $P_\mathrm{oh}$ might be sensitive to the magnitude of $\eta$.

We extract the X-type grids with 2D extrema of $J_\parallel$ at $t=9$ to compare its distribution with that of all X-type grids (Fig.\,\ref{fig:hist_x_180}).
The number densities of $\left|J_\parallel\right|$ and $P_\mathrm{oh}$ of 2D extreme grids almost coincide with that of all X-grids for large values (see the $\left|J_\parallel\right|>300$ region in Fig.\,\ref{fig:hist_x_180}a and the $P_\mathrm{oh}>0.6$ region in Fig.\,\ref{fig:hist_x_180}b).
But for smaller values, they are lower than the curves of all X-grids (see Fig.\,\ref{fig:hist_x_180}a and b).
For $\theta_\mathrm{eig}$ (Fig.\,\ref{fig:hist_x_180}c), compared with all X-type grids, the number density curve of 2D extreme grids has a similar shape but is about one order lower.
The two curves of $R_\mathrm{tr}$ have similar trends (Fig.\,\ref{fig:hist_x_180}d), but 2D extreme grids tend to concentrate within a smaller value region ($R_\mathrm{tr}<0.1$), implying that, near the extremal-$\Xi$ lines, the variations of local field strength on the direction of guide field ($\hat{\mathbf{e}}_3$) is weaker.

The statistical results of 3D O-type grids are also provided in Appendix \ref{asec:OGrids}.
The rules of $\left|J_\parallel\right|$, $P_\mathrm{oh}$, and $R_\mathrm{tr}$ of 3D O-type grids are similar to 3D X-type grids (see Figs.\ref{fig:hist_o_alltime} and \ref{fig:hist_o_180}).
The main difference is that more O-type grids with large $\theta_\mathrm{eig}$ near $90^\circ$ are identified, corresponding to flux ropes with poloidal field lines close to circle shapes (see Figs.\ref{fig:hist_o_alltime}c and \ref{fig:hist_o_180}c).

\begin{figure*}
\centering
\includegraphics[width=1\textwidth]{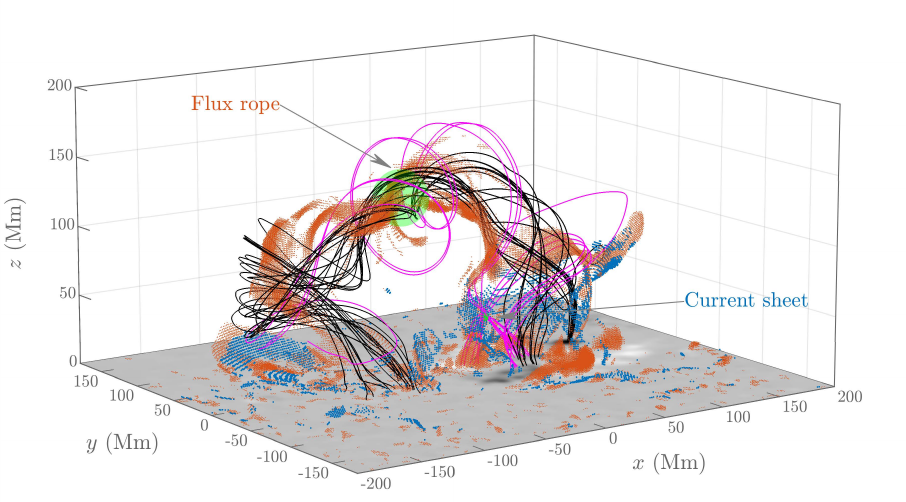}
\caption{Distributions of 3D X-type (blue dots) and O-type (orange dots) grids in a data-driven simulation of a large-scale coronal flux rope eruption.
The plotted grids are 2D projected extreme points of $J_\parallel$, satisfying $\left|J_\parallel\right|>11.9\,\mathrm{statC\,s^{-1}\,cm^{-2}}$.
The black curves are field lines traced from the initial sampling positions near the top of the flux rope as marked by the big green sphere, while the magenta ones are traced from a small region in the eruption current sheet as marked by the small green sphere.
The distribution of $B_z$ is overlapped at the bottom.
}
\label{fig:DD}
\end{figure*}

\section{Benchmark 2: Large-scale Magnetic Structures during a Coronal Flux Rope Eruption\label{sec:test2}}

Besides the statistical analysis of local reconnection structures, our method can also be applied to recognize large-scale magnetic structures within 3D simulations.
Here we use the data from a data-driven simulation of an erupting coronal flux rope by \cite{Guo2023a} to test our method.
\cite{Guo2023a} solved the full MHD equations in which a realistic coronal physical environment is considered.
The simulation initially set a flux rope comparable with observations and proceeded under the constraint of a bottom condition obtained from observed magnetic and velocity fields.
The simulation reproduced the observational characteristics of the X1.0 flare on 2021 October 28.
Detailed simulation configurations are introduced in \cite{Guo2023a}.

We apply our method to analyze the magnetic field data at 15:32 UT.
At this moment, a major flux rope is rising and an eruption current sheet forms under the flux rope \citep[see][Fig.\,3d]{Guo2023a}.
As shown by orange dots and black field lines in Fig.\,\ref{fig:DD}, the O-type (types 2 and 3) grids outline the position of the major flux rope, which is consistent with the results of \cite{Guo2023a}.
Our method also precisely locates the position of the eruption current sheet (see Fig.\,\ref{fig:DD}), wherein sheared field lines form a current sheet that plays an important role in driving the eruption of the flux rope.
If we trace a bundle of field lines starting from a local region in the current sheet, these field lines show a locally sheared pattern and then extend to the outer layer of the flux rope (see the magenta curves in Fig.\,\ref{fig:DD}).

\begin{figure*}
\centering
\includegraphics[width=1\textwidth]{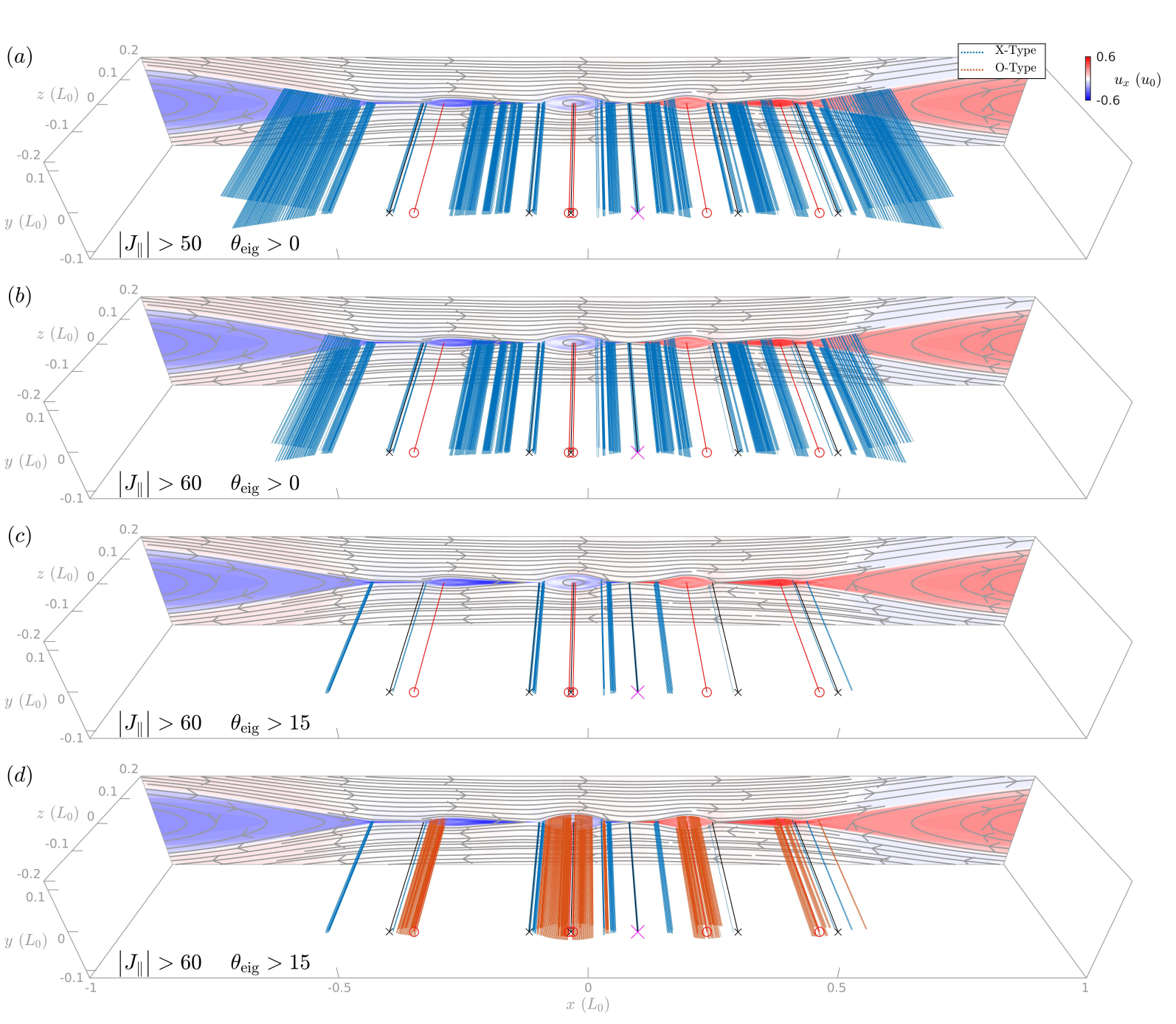}
\caption{The distributions of 3D X and O-type grids within a 2.5D Harris-sheet reconnection.
The 3D magnetic field has $z$-symmetry and is constructed from a 2.5D simulation with the same configurations as the 3D Harris sheet in Sect.\,\ref{sec:test1}. 
This figure shows the result at $t=4$.
In panel (a), the blue dots have the same definition as in Figs.\,\ref{fig:hs3d} and \ref{fig:hs3d_ev}.
The black ``x'' and red ``o'' markers depict the positions of 2D X and O points on the $x$-$y$ plane as determined by the 2D method, while the black and red lines are the corresponding X and O lines in 3D.
The principal reconnection site with strongest $\left|J_\parallel\right|$ is marked by a magenta ``x'' marker.
On the plane $z=0.2$, the profile of $u_x$ is depicted while the magnetic field lines are shown by gray curves.
Panels (b) and (c) are similar to panel (a) but exhibit results with different thresholds of $\left|J_\parallel\right|$ and $\theta_\mathrm{eig}$.
Panel (d) has the same thresholds as (c) but also plots 3D O-type grids with orange dots.
}
\label{fig:hs2d}
\end{figure*}

\begin{figure*}
\centering
\includegraphics[width=0.6\textwidth]{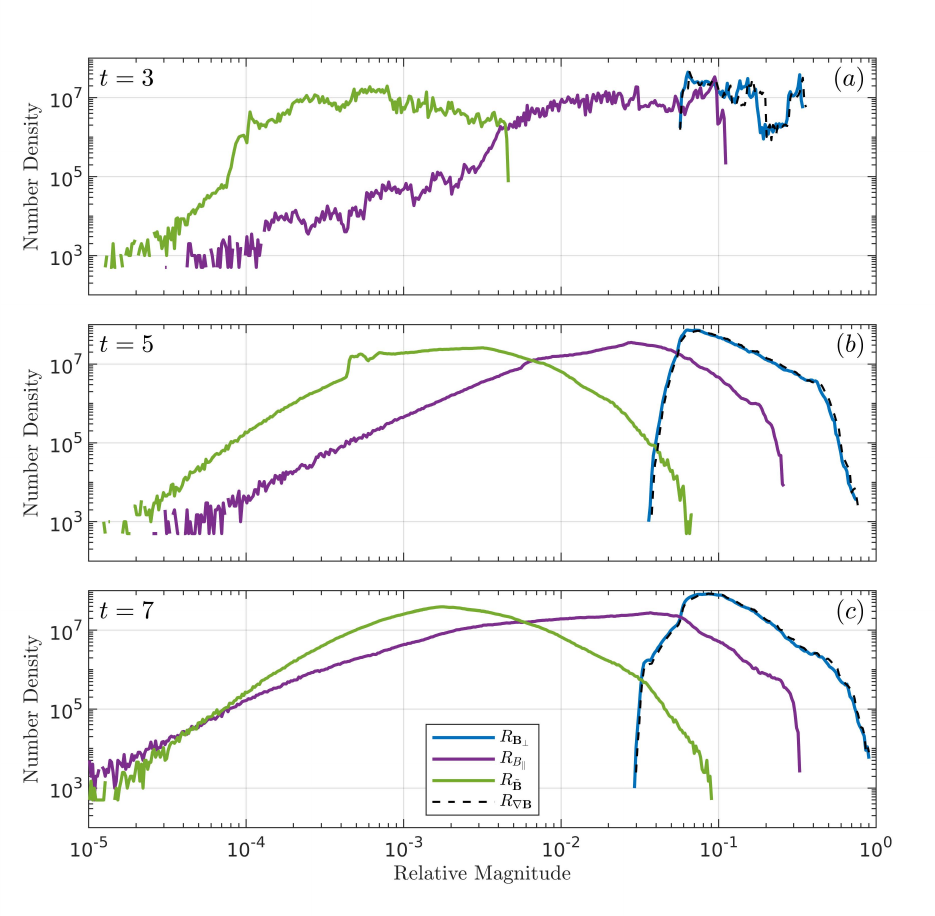}
\caption{The number densities of $R_{\mathbf{B}_\perp}$ (blue curves), $R_{B_\parallel}$ (purple curves), $R_{\tilde{\mathbf{B}}}$ (green curves), and $R_{\nabla\mathbf{B}}$ (black dashed curves) of all grids with $\left|J_\parallel\right|>50$ at different moments in the Harris-sheet simulation.
}
\label{fig:mppapprox}
\end{figure*}

\section{Summary and Discussion\label{sec:conclude}}

To conclude, we develop an efficient method for analyzing 3D magnetic reconnection from discrete simulation data.
Based on the general reconnection theory by \cite{Hesse1988}, we discuss the usage of the nonideal electric field $E_\parallel$ in representing the reconnection sites and propose that the locations of $\nabla_\perp E_\parallel=0$ can present the distribution of extremal-$\Xi$ lines.
We perform theoretical analysis on the local magnetic structure and provide a complete classification including nine types of magnetic structures (Table\,\ref{tab:types}).
By defining two parameters, $\theta_\mathrm{eig}$ and $R_\mathrm{tr}$, the geometric properties of the local magnetic field can be further clarified.
We construct an efficient numerical method, which only performs algebraical manipulations on the discrete magnetic field, avoiding computationally expensive operations like field-line tracing and root-finding. 
This method directly outputs the classification and geometric parameters of local magnetic fields at arbitrary grids, which has been implemented in \textsf{Matlab} language and can be freely obtained on GitHub.

Through two typical numerical benchmarks, namely the 3D Harris-sheet turbulent reconnection and the coronal flux rope eruption, we show that this method can precisely identify the local structures of discrete magnetic field (Fig.\,\ref{fig:diagram}).
Through the strength of the nonideal electric field and the geometric attributes of magnetic field, the local structures of reconnection sites can be effectively revealed.
We exhibit not only qualitative pictures but also quantitative statistical results of the 3D turbulent reconnection by use of the outputs of our method.
It is also shown that macro-scale magnetic structures such as flux ropes and eruption current sheets can be recognized by our method.

As a generalization of the 2D method for locating reconnection sites, our method can also determine the classic 2D X/O lines.
To clarify this, we run a 2.5D simulation of the Harris-sheet reconnection in Sect.\,\ref{sec:test1} and copy the simulation data along $z$-grids to construct 3D data with translation symmetry on $z$-direction.
From the perspective of 2D methods, one naturally chooses the $z$-direction as the ``guide field'' direction and the $x$-$y$ plane as a ``global'' projection plane, on which the 2D null points with X or O-type structures can be located \citep[see Fig.\,\ref{fig:hs2d} and also see][]{Huang2010,Shen2011,Lynch2016,Ye2019,Wang2021b}.
By using our 3D method, the 3D X and O-type grids can also be located (Fig.\,\ref{fig:hs2d}).
Because the fixed global projection plane in the 2D method has been generalized as the spatially variable local MPPs, more X-type points can be recognized.
After imposing the constraint of 2D extreme $J_\parallel$ and further selecting grids with larger $\left|J_\parallel\right|$ and $\theta_\mathrm{eig}$, we find that the resultant distribution of 3D X-grids gets almost consistent with the 2D X-lines (see Fig.\,\ref{fig:hs2d}a--c).
Especially, the X-line of the principal reconnection site can be well resolved (see the magenta ``x'' markers in Fig.\,\ref{fig:hs2d}).
Moreover, the 3D O-grids recognized by our method outline a flux-rope volume full of twisted field lines, while the 2D method can only locate its axis (O-line) defined on the $x$-$y$ plane. 

Different from the global analysis techniques requiring field-line tracing, our method performs local analysis on discrete grids.
Despite the promising performances in determining global magnetic structures, field-line tracing techniques can hardly be applied to the analysis of massive discrete data of turbulence mainly for two reasons.
First, unlike the laminar parts of a magnetic field, the turbulent parts have no clear global topological patterns but are full of multi-scale chaotic structures.
Second, turbulent structures with smaller scales correspond to larger local gradients which are more sensitive to numerical truncation errors.
The numerical integration for tracing field lines can further amplify the influences of local errors by global iterations, which makes the results unreliable.

The technique proposed by \cite{Lapenta2021} is also a local method, which defines a local frame based on the local electric field $\mathbf{E}$ and electric drift speed.
Because the Lorentz boost speed to eliminate the magnetic field perpendicular to $\bf E$, namely $\mathbf{v}_{L}=c^2\bf{E\times B/E^2}$, can only be lower than the speed of light $c$ in the adjacent region of reconnection sites, \cite{Lapenta2021} suggested that 3D reconnection sites can be located by finding locations with $v_L<c$, which has been verified within kinetic simulation data \citep{Lapenta2023a}.
But for MHD simulations, the grid sizes, typically several orders larger than the inertial scales of electrons and ions, cannot resolve the absolute value of electric field within the dissipation regions.
As a result, when we applied this method to MHD data, no reconnection site with $v_L<c$ can be recognized.
In contrast, our method mainly analyzes the magnetic field and only uses the relative strength of nonideal electric field as an auxiliary parameter, which is thus independent of the system scales.
Therefore, besides MHD simulations, our method can also be potentially applied to kinetic simulations, plasma experiments, and even \emph{in-situ} observations. 

Now we discuss the limitations of our method.
First, when analyzing local magnetic structures, our method depends on a finite magnetic field to establish a local frame, which gets invalid at magnetic null points.
For simulation data, the magnetic strength on grids can hardly be zeros, which makes it safe to apply our method.
To obtain complete information on 3D reconnection sites, one still needs to run a 3D null-point analyzing program, which has been implemented as the \texttt{ANP} function in the \textsf{LoRD} toolkit.
However, because 3D reconnection takes place in the vicinity of a null point (not just at the null point) where current density and $E_\parallel$ concentrate \citep[see][]{Pontin2007}, our method is still useful near null points. 

Second, when analyzing the local magnetic structures near a grid $\mathbf{r}^{lmn}$, we approximate the surface normal to magnetic field by a plane (MPP) within a finite volume with the length scale of $\Delta L$, which validates better if $R_{\nabla\mathbf{B}}\equiv\Delta L\left\Vert\mathbf{D}'^{lmn}_B\right\Vert/\left|\mathbf{B}^{lmn}\right|\ll 1$.
Meanwhile, we mainly use the perpendicular magnetic field on the MPP ($\mathbf{B}_\perp$) to determine the local magnetic structures which might also be affected by $B_\parallel$ and $\tilde{\mathbf{B}}$.
In applications, we define three ratios to evaluate the importance of the three terms, namely, $R_{\mathbf{B}_\perp}\equiv\Delta L\left\Vert\mathbf{M}^{lmn}\right\Vert/\left|\mathbf{B}^{lmn}\right|$, $R_{B_\parallel}\equiv\Delta L\sqrt{\left(D'^{lmn}_{B,31}\right)^2+\left(D'^{lmn}_{B,32}\right)^2}/\left|\mathbf{B}^{lmn}\right|$, and $R_{\tilde{\mathbf{B}}}\equiv\Delta L\sqrt{\left(D'^{lmn}_{B,13}\right)^2+\left(D'^{lmn}_{B,23}\right)^2+\left(D'^{lmn}_{B,33}\right)^2}/\left|\mathbf{B}^{lmn}\right|$.
In the Harris-sheet simulation, for most grids, the magnitude of $R_{\nabla\mathbf{B}}$ is $\sim 0.1$ which implies the MPP approximation is acceptable (see the black dashed curves in Fig.\,\ref{fig:mppapprox}). 
Moreover, the distribution of $R_{\mathbf{B}_\perp}$ almost overlaps with that of $R_{\nabla\mathbf{B}}$ (see the blue curves in Fig.\,\ref{fig:mppapprox}), the peak of $R_{B_\parallel}$ is about one order smaller than $R_{\mathbf{B}_\perp}$, and the magnitude of $R_{\tilde{\mathbf{B}}}$ is much smaller even compared with $R_{B_\parallel}$.
Therefore, in this simulation, the variation of magnetic field in a cell is dominated by $\mathbf{B}_\perp$ for most reconnection grids.
$R_{\nabla\mathbf{B}}$, $R_{\mathbf{B}_\perp}$, $R_{B_\parallel}$, and $R_{\tilde{\mathbf{B}}}$ can be directly calculated by the 1--10 columns of \texttt{RDInfo.ExtraData} (see Table \ref{tab:lordextradata}), which is helpful for users to evaluate the validations of MPP approximation and the identified local magnetic structures.
We have not added the four ratios to the output of \texttt{ARD} for the sake of saving storage space.

Third, our method provides the reconnection properties of the finest objects in discrete data (i.e., the grids), which, however, cannot determine the global properties of small-scale reconnection regions containing a series of adjacent grids.
Based on the information of grids, it is possible to further recognize local finite-volume reconnection regions through clumping algorithms, which might produce more useful information like volume-integrated parameters and topological properties.
We will attempt to add these functions in the future to make the code more powerful in dealing with multi-scale reconnection processes.

\begin{acknowledgements}
The authors would like to thank the anonymous referee for constructive suggestions and acknowledge J. Chen, P. Fan, and C. Xing for their helpful discussions.
The simulations in this paper were performed in the cluster system of the High Performance Computing Center (HPCC) of Nanjing University.
This research is supported by the Strategic Priority Research Program of the Chinese Academy of Sciences (Grant No. XDB0560000), by the National Key R\&D Program of China (Grant No. 2021YFA1600504), and by the Natural Science Foundation of China (Grant No. 12127901).
\end{acknowledgements}

\bibliographystyle{aa}
\bibliography{library}

\begin{thebibliography}{60}
\expandafter\ifx\csname natexlab\endcsname\relax\def\natexlab#1{#1}\fi

\bibitem[{Aulanier {et~al.}(2006)Aulanier, Pariat, D{\'{e}}moulin, \& Devore}]{Aulanier2006}
Aulanier, G., Pariat, E., D{\'{e}}moulin, P., \& Devore, C.~R. 2006, Sol. Phys., 238, 347

\bibitem[{Comisso \& Sironi(2022)}]{Comisso2022}
Comisso, L. \& Sironi, L. 2022, Astrophys. J. Lett., 936, L27

\bibitem[{Demoulin {et~al.}(1997)Demoulin, Bagala, Mandrini, Henoux, \& Rovira}]{Demoulin1997}
Demoulin, P., Bagala, L.~G., Mandrini, C.~H., Henoux, J.~C., \& Rovira, M.~G. 1997, Astron. Astrophys., 325, 305

\bibitem[{Demoulin {et~al.}(1996)Demoulin, Henoux, Priest, \& Mandrini}]{Demoulin1996a}
Demoulin, P., Henoux, J.~C., Priest, E.~R., \& Mandrini, C.~H. 1996, Astron. Astrophys., 308, 643

\bibitem[{D{\'{e}}moulin {et~al.}(1996)D{\'{e}}moulin, Priest, \& Lonie}]{Demoulin1996}
D{\'{e}}moulin, P., Priest, E.~R., \& Lonie, D.~P. 1996, J. Geophys. Res. Sp. Phys., 101, 7631

\bibitem[{Dong {et~al.}(2022)Dong, Wang, Huang, Comisso, Sandstrom, \& Bhattacharjee}]{Dong2022}
Dong, C., Wang, L., Huang, Y.-M., {et~al.} 2022, Sci. Adv., 8, eabn7627

\bibitem[{Fu {et~al.}(2015)Fu, Vaivads, Khotyaintsev, Olshevsky, Andr{\'{e}}, Cao, Huang, Retin{\`{o}}, \& Lapenta}]{Fu2015}
Fu, H.~S., Vaivads, A., Khotyaintsev, Y.~V., {et~al.} 2015, J. Geophys. Res. Sp. Phys., 120, 3758

\bibitem[{Greco {et~al.}(2017)Greco, Matthaeus, Perri, Osman, Servidio, Wan, \& Dmitruk}]{Greco2017}
Greco, A., Matthaeus, W.~H., Perri, S., {et~al.} 2017, Space Sci. Rev., 214, 1

\bibitem[{Guo {et~al.}(2023)Guo, Ni, Zhong, Guo, Xia, Li, Poedts, Schmieder, \& Chen}]{Guo2023a}
Guo, J.~H., Ni, Y.~W., Zhong, Z., {et~al.} 2023, Astrophys. J. Suppl. Ser., 266, 3

\bibitem[{Hada {et~al.}(2003)Hada, Koga, \& Yamamoto}]{Hada2003}
Hada, T., Koga, D., \& Yamamoto, E. 2003, Space Sci. Rev., 107, 463

\bibitem[{Haynes \& Parnell(2007)}]{Haynes2007}
Haynes, A.~L. \& Parnell, C.~E. 2007, Phys. Plasmas, 14, 82107

\bibitem[{Haynes \& Parnell(2010)}]{Haynes2010}
Haynes, A.~L. \& Parnell, C.~E. 2010, Phys. Plasmas, 17, 92903

\bibitem[{Hesse {et~al.}(2005)Hesse, Forbes, \& Birn}]{Hesse2005}
Hesse, M., Forbes, T.~G., \& Birn, J. 2005, Astrophys. J., 631, 1227

\bibitem[{Hesse \& Schindler(1988)}]{Hesse1988}
Hesse, M. \& Schindler, K. 1988, J. Geophys. Res. Sp. Phys., 93, 5559

\bibitem[{Huang \& Bhattacharjee(2010)}]{Huang2010}
Huang, Y.~M. \& Bhattacharjee, A. 2010, Phys. Plasmas, 17, 062104

\bibitem[{Huang \& Bhattacharjee(2016)}]{Huang2016}
Huang, Y.-M. \& Bhattacharjee, A. 2016, Astrophys. J., 818, 20

\bibitem[{Isliker {et~al.}(2019)Isliker, Archontis, \& Vlahos}]{Isliker2019}
Isliker, H., Archontis, V., \& Vlahos, L. 2019, Astrophys. J., 882, 57

\bibitem[{Ji {et~al.}(2022)Ji, Daughton, Jara-Almonte, Le, Stanier, \& Yoo}]{Ji2022}
Ji, H., Daughton, W., Jara-Almonte, J., {et~al.} 2022, Nat. Rev. Phys., 4, 263

\bibitem[{Komar {et~al.}(2013)Komar, Cassak, Dorelli, Glocer, \& Kuznetsova}]{Komar2013}
Komar, C.~M., Cassak, P.~A., Dorelli, J.~C., Glocer, A., \& Kuznetsova, M.~M. 2013, J. Geophys. Res. Sp. Phys., 118, 4998

\bibitem[{Kowal {et~al.}(2020)Kowal, Falceta-Gon{\c{c}}alves, Lazarian, \& Vishniac}]{Kowal2020}
Kowal, G., Falceta-Gon{\c{c}}alves, D.~A., Lazarian, A., \& Vishniac, E.~T. 2020, Astrophys. J., 892, 50

\bibitem[{Lapenta(2021)}]{Lapenta2021}
Lapenta, G. 2021, Astrophys. J., 911, 147

\bibitem[{Lapenta(2023)}]{Lapenta2023a}
Lapenta, G. 2023, Nat. Phys., 19, 159

\bibitem[{Lau \& Finn(1990)}]{Lau1990}
Lau, Y.-T. \& Finn, J.~M. 1990, Astrophys. J., 350, 672

\bibitem[{Li {et~al.}(2022)Li, Cheng, Guo, Yan, Wang, Zhong, Li, \& Ding}]{Li2022}
Li, H.~T., Cheng, X., Guo, J.~H., {et~al.} 2022, Astron. Astrophys., 663

\bibitem[{Li {et~al.}(2021{\natexlab{a}})Li, Priest, \& Guo}]{Li2021a}
Li, T., Priest, E., \& Guo, R. 2021{\natexlab{a}}, Proc. R. Soc. A Math. Phys. Eng. Sci., 477, 20200949

\bibitem[{Li {et~al.}(2021{\natexlab{b}})Li, Liu, \& Qi}]{Li2021b}
Li, T.~C., Liu, Y.-H., \& Qi, Y. 2021{\natexlab{b}}, Astrophys. J. Lett., 909, L28

\bibitem[{Li {et~al.}(2023)Li, Liu, Qi, \& Zhou}]{Li2023b}
Li, T.~C., Liu, Y.-H., Qi, Y., \& Zhou, M. 2023, Phys. Rev. Lett., 131, 85201

\bibitem[{Lynch {et~al.}(2016)Lynch, Edmondson, Kazachenko, \& Guidoni}]{Lynch2016}
Lynch, B.~J., Edmondson, J.~K., Kazachenko, M.~D., \& Guidoni, S.~E. 2016, Astrophys. J., 826, 43

\bibitem[{Meyer {et~al.}(2014)Meyer, Balsara, \& Aslam}]{Meyer2014}
Meyer, C.~D., Balsara, D.~S., \& Aslam, T.~D. 2014, J. Comput. Phys., 257, 594

\bibitem[{Miyoshi \& Kusano(2005)}]{Miyoshi2005}
Miyoshi, T. \& Kusano, K. 2005, J. Comput. Phys., 208, 315

\bibitem[{Olshevsky {et~al.}(2020)Olshevsky, Pontin, Williams, Parnell, Fu, Liu, Yao, \& Khotyaintsev}]{Olshevsky2020}
Olshevsky, V., Pontin, D.~I., Williams, B., {et~al.} 2020, Astron. Astrophys., 644

\bibitem[{Parnell {et~al.}(2010{\natexlab{a}})Parnell, Haynes, \& Galsgaard}]{Parnell2010}
Parnell, C.~E., Haynes, A.~L., \& Galsgaard, K. 2010{\natexlab{a}}, J. Geophys. Res. Sp. Phys., 115

\bibitem[{Parnell {et~al.}(2010{\natexlab{b}})Parnell, Maclean, \& Haynes}]{Parnell2010a}
Parnell, C.~E., Maclean, R.~C., \& Haynes, A.~L. 2010{\natexlab{b}}, Astrophys. J. Lett., 725, L214

\bibitem[{Parnell {et~al.}(1996)Parnell, Smith, Neukirch, \& Priest}]{Parnell1996}
Parnell, C.~E., Smith, J.~M., Neukirch, T., \& Priest, E.~R. 1996, Phys. Plasmas, 3, 759

\bibitem[{Pontin {et~al.}(2007)Pontin, Bhattacharjee, \& Galsgaard}]{Pontin2007}
Pontin, D.~I., Bhattacharjee, A., \& Galsgaard, K. 2007, Phys. Plasmas, 14, 52106

\bibitem[{Pontin \& Priest(2022)}]{Pontin2022}
Pontin, D.~I. \& Priest, E.~R. 2022, Living Rev. Sol. Phys., 19, 1

\bibitem[{Priest \& Forbes(2000)}]{Priest2000}
Priest, E. \& Forbes, T. 2000, {Magnetic Reconnection: MHD Theory and Applications} (Cambridge: Cambridge University Press)

\bibitem[{Priest \& D{\'{e}}moulin(1995)}]{Priest1995}
Priest, E.~R. \& D{\'{e}}moulin, P. 1995, J. Geophys. Res. Sp. Phys., 100, 23443

\bibitem[{Priest \& Forbes(1989)}]{Priest1989}
Priest, E.~R. \& Forbes, T.~G. 1989, Sol. Phys., 119, 211

\bibitem[{Priest {et~al.}(2003)Priest, Hornig, \& Pontin}]{Priest2003}
Priest, E.~R., Hornig, G., \& Pontin, D.~I. 2003, J. Geophys. Res. Sp. Phys., 108

\bibitem[{Reid {et~al.}(2020)Reid, Parnell, Hood, \& Browning}]{Reid2020}
Reid, J., Parnell, C.~E., Hood, A.~W., \& Browning, P.~K. 2020, Astron. Astrophys., 633, A92

\bibitem[{Schindler {et~al.}(1988)Schindler, Hesse, \& Birn}]{Schindler1988}
Schindler, K., Hesse, M., \& Birn, J. 1988, J. Geophys. Res. Sp. Phys., 93, 5547

\bibitem[{Shen {et~al.}(2011)Shen, Lin, \& Murphy}]{Shen2011}
Shen, C., Lin, J., \& Murphy, N.~A. 2011, Astrophys. J., 737, 14

\bibitem[{Stone {et~al.}(2020)Stone, Tomida, White, \& Felker}]{Stone2020}
Stone, J.~M., Tomida, K., White, C.~J., \& Felker, K.~G. 2020, Astrophys. J. Suppl. Ser., 249, 4

\bibitem[{Titov(2007)}]{Titov2007}
Titov, V.~S. 2007, Astrophys. J., 660, 863

\bibitem[{Titov {et~al.}(2009)Titov, Forbes, Priest, Miki{\'{c}}, \& Linker}]{Titov2009}
Titov, V.~S., Forbes, T.~G., Priest, E.~R., Miki{\'{c}}, Z., \& Linker, J.~A. 2009, Astrophys. J., 693, 1029

\bibitem[{Titov {et~al.}(2002)Titov, Hornig, \& D{\'{e}}moulin}]{Titov2002}
Titov, V.~S., Hornig, G., \& D{\'{e}}moulin, P. 2002, J. Geophys. Res. Sp. Phys., 107, SSH 3

\bibitem[{Vlahos \& Isliker(2023)}]{Vlahos2023}
Vlahos, L. \& Isliker, H. 2023, Phys. Plasmas, 30, 40502

\bibitem[{Wang {et~al.}(2023)Wang, Cheng, Ding, Liu, Liu, \& Zhu}]{Wang2023a}
Wang, Y., Cheng, X., Ding, M., {et~al.} 2023, Astrophys. J. Lett., 954, L36

\bibitem[{Wang {et~al.}(2021)Wang, Cheng, Ding, \& Lu}]{Wang2021b}
Wang, Y., Cheng, X., Ding, M., \& Lu, Q. 2021, Astrophys. J., 923, 227

\bibitem[{Wilmot-Smith \& Priest(2007)}]{Wilmot-Smith2007}
Wilmot-Smith, A.~L. \& Priest, E.~R. 2007, Phys. Plasmas, 14, 102903

\bibitem[{Wyper \& Hesse(2015)}]{Wyper2015}
Wyper, P.~F. \& Hesse, M. 2015, Phys. Plasmas, 22, 42117

\bibitem[{Ye {et~al.}(2020)Ye, Cai, Shen, Raymond, Lin, Roussev, \& Mei}]{Ye2020}
Ye, J., Cai, Q., Shen, C., {et~al.} 2020, Astrophys. J., 897, 64

\bibitem[{Ye {et~al.}(2023)Ye, Raymond, Mei, Cai, Chen, Li, \& Lin}]{Ye2023a}
Ye, J., Raymond, J.~C., Mei, Z., {et~al.} 2023, Astrophys. J., 955, 88

\bibitem[{Ye {et~al.}(2019)Ye, Shen, Raymond, Lin, \& Ziegler}]{Ye2019}
Ye, J., Shen, C., Raymond, J.~C., Lin, J., \& Ziegler, U. 2019, Mon. Not. R. Astron. Soc., 482, 588

\bibitem[{Zhang {et~al.}(2022)Zhang, Chen, Liu, \& Wang}]{Zhang2022}
Zhang, P., Chen, J., Liu, R., \& Wang, C. 2022, Astrophys. J., 937, 26

\bibitem[{Zhang {et~al.}(2021)Zhang, Guo, Daughton, Li, \& Li}]{Zhang2021}
Zhang, Q., Guo, F., Daughton, W., Li, H., \& Li, X. 2021, Phys. Rev. Lett., 127, 185101

\bibitem[{Zhang {et~al.}(2023)Zhang, Fu, Cao, Wang, \& Liu}]{Zhang2023}
Zhang, W.~Z., Fu, H.~S., Cao, J.~B., Wang, Z., \& Liu, Y.~Y. 2023, Astrophys. J., 953, 23

\bibitem[{Zhdankin {et~al.}(2013)Zhdankin, Uzdensky, Perez, \& Boldyrev}]{Zhdankin2013}
Zhdankin, V., Uzdensky, D.~A., Perez, J.~C., \& Boldyrev, S. 2013, Astrophys. J., 771, 124

\bibitem[{Zhong {et~al.}(2021)Zhong, Guo, \& Ding}]{Zhong2021}
Zhong, Z., Guo, Y., \& Ding, M.~D. 2021, Nat. Commun., 12, 1

\end{thebibliography}

\begin{appendix}

\section{Analysis of Local Reconnection Effects\label{asec:localeffects}}

To quantitatively compare the $E_\parallel$ and $\mathbf{N}_\perp$ terms in Eq.\,\ref{eq:GMRLocal}, here we deduce a different form of this equation.
The basic idea is transforming the RHS of Eq.\,\ref{eq:GMRLocal} from the flux coordinate frame to the local orthogonal and normalized frame spanned by $\hat{\bf e}_1$, $\hat{\bf e}_2$, and $\hat{\bf e}_3$.
In this section, we use $\mathbf{x}=\left(x^1,x^2,x^3\right)^T$ and $\mathbf{x}'=\left(x'^1,x'^2,x'^3\right)\equiv\left(\alpha,\beta,s\right)^T$ to denote the coordinates in the $\left(\hat{\bf e}_1,\hat{\bf e}_2,\hat{\bf e}_3\right)$ and $\left(\nabla\alpha,\nabla\beta,\hat{\bf b}\right)$ frames, respectively.
Notice that $\hat{\bf e}_3=\hat{\bf b}$, $x^3=x'^3=s$, and $\partial/\partial s=\partial/\partial x^3$.

First, based on the vector relations
\begin{align}
   x^1\hat{\bf e}_1 & +x^2\hat{\bf e}_2=\alpha\nabla\alpha+\beta\nabla\beta\,,\label{eq:a_vtran}\\
   \nabla\alpha & =\left(\hat{\bf e}_1\cdot\nabla\alpha\right)\hat{\bf e}_1+\left(\hat{\bf e}_2\cdot\nabla\alpha\right)\hat{\bf e}_2\,,\label{eq:a_alpha}\\
   \nabla\beta & =\left(\hat{\bf e}_1\cdot\nabla\beta\right)\hat{\bf e}_1+\left(\hat{\bf e}_2\cdot\nabla\beta\right)\hat{\bf e}_2\,,\label{eq:a_beta}
\end{align}
the transformation matrices between the two frames can be obtained as
\begin{align}
   \frac{\partial\mathbf{x}}{\partial\mathbf{x}'} & = \left(\begin{array}{ccc}
      \hat{\bf e}_1\cdot\nabla\alpha,& \hat{\bf e}_1\cdot\nabla\beta,&0\\
      \hat{\bf e}_2\cdot\nabla\alpha,& \hat{\bf e}_2\cdot\nabla\beta,&0\\
      0,& 0,& 1\\
   \end{array}\right)\,,\label{eq:a_Tinv}\\
   \frac{\partial\mathbf{x}'}{\partial\mathbf{x}} & = \frac{1}{B}\left(\begin{array}{ccc}
      \hat{\bf e}_2\cdot\nabla\beta, & -\hat{\bf e}_1\cdot\nabla\beta,&0\\
      -\hat{\bf e}_2\cdot\nabla\alpha,& \hat{\bf e}_1\cdot\nabla\alpha,&0\\
      0,& 0,& B\\
   \end{array}\right)\,,\label{eq:a_T}
\end{align}
where $B=\hat{\bf b}\cdot\left(\nabla\alpha\times\nabla\beta\right)$ is the magnetic strength at the origin point.

Second, we rewrite the $\partial E_\parallel/\partial\beta$ and $\partial E_\parallel/\partial\alpha$ terms in Eq.\,\ref{eq:GMRLocal}.
Considering that $\mathbf{E}\cdot\hat{\bf b}=\mathbf{E}\cdot\hat{\bf e}_3$, $E_\parallel$ is invariant in the new frame.
According to Eq.\,\ref{eq:a_Tinv}, we have $\partial/\partial\alpha=\left(\hat{\bf e}_1\cdot\nabla\alpha\right)\partial/\partial x^1+\left(\hat{\bf e}_2\cdot\nabla\alpha\right)\partial/\partial x^2$ and $\partial/\partial\beta=\left(\hat{\bf e}_1\cdot\nabla\beta\right)\partial/\partial x^1+\left(\hat{\bf e}_2\cdot\nabla\beta\right)\partial/\partial x^2$, and thus
\begin{align}
   \frac{\partial E_\parallel}{\partial\alpha} & = \frac{\partial E_\parallel}{\partial x^1}\hat{\bf e}_1\cdot\nabla\alpha+\frac{\partial E_\parallel}{\partial x^2}\hat{\bf e}_2\cdot\nabla\alpha = \nabla E_\parallel\cdot\nabla\alpha\,,\\
   \frac{\partial E_\parallel}{\partial\beta} & = \frac{\partial E_\parallel}{\partial x^1}\hat{\bf e}_1\cdot\nabla\beta+\frac{\partial E_\parallel}{\partial x^2}\hat{\bf e}_2\cdot\nabla\beta = \nabla E_\parallel\cdot\nabla\beta\,.
\end{align}

Third, we rewrite the $\partial N^\beta/\partial s$ and $\partial N^\alpha/\partial s$ terms in Eq.\,\ref{eq:GMRLocal}.
According to Eq.\,\ref{eq:a_T}, we have
\begin{align}
   N^\alpha & = \frac{1}{B}\left(N^1\hat{\bf e}_2\cdot\nabla\beta-N^2\hat{\bf e}_1\cdot\nabla\beta\right)\,,\label{eq:a_N1}\\
   N^\beta& = -\frac{1}{B}\left(N^1\hat{\bf e}_2\cdot\nabla\alpha-N^2\hat{\bf e}_1\cdot\nabla\alpha\right)\,.\label{eq:a_N2}
\end{align}
Therefore, 
\begin{align}
   \frac{\partial N^\alpha}{\partial s} = & \left[\hat{\bf e}_2\frac{\partial}{\partial x^3}\left(\frac{N^1}{B}\right)-\hat{\bf e}_1\frac{\partial}{\partial x^3}\left(\frac{N^2}{B}\right)\right]\cdot\nabla\beta \nonumber\\
   & +\left(\frac{N^1}{B}\frac{\partial\hat{\bf e}_2}{\partial x^3} - \frac{N^2}{B}\frac{\partial\hat{\bf e}_1}{\partial x^3}\right)\cdot\nabla\beta \nonumber\\
   & +\frac{1}{B}\left(N^1\hat{\bf e}_2-N^2\hat{\bf e}_1\right)\cdot\frac{\partial\nabla\beta}{\partial x^3}\,.\label{eq:a_NaDs}
\end{align}
Using $\nabla\left(\hat{\bf b}\cdot\nabla\beta\right)=0$, it can be proved that $\partial\nabla\beta/\partial x^3=-\nabla\hat{\bf b}\cdot\nabla\beta=-\nabla\hat{\bf e}_3\cdot\nabla\beta$ and similarly $\partial\nabla\alpha/\partial x^3=-\nabla\hat{\bf e}_3\cdot\nabla\alpha$.
As a result, we can define a vector
\begin{align}
   {\bf \Gamma}=& \hat{\bf e}_2\frac{\partial\left(N^1/B\right)}{\partial x^3}-\hat{\bf e}_1\frac{\partial\left(N^2/B\right)}{\partial x^3}+\frac{N^1}{B}\frac{\partial\hat{\bf e}_2}{\partial x^3}\nonumber \\
   & - \frac{N^2}{B}\frac{\partial\hat{\bf e}_1}{\partial x^3}+ \left(-\frac{N^1}{B}\hat{\bf e}_2+\frac{N^2}{B}\hat{\bf e}_1\right)\cdot\nabla\hat{\bf e}_3\,,\label{eq:a_Gamma}
\end{align}
and write Eq.\,\ref{eq:a_NaDs} as $\partial N^\alpha/\partial s={\bf \Gamma}\cdot\nabla\beta$.
Similarly, we have $\partial N^\beta/\partial s=-{\bf \Gamma}\cdot\nabla\alpha$.

Finally, Eq.\,\ref{eq:GMRLocal} becomes
\begin{subequations}\label{eq:a_GMRLocalNew}
\begin{align}
   \frac{\partial\dot{\alpha}}{\partial x^3} & = \nabla E_\parallel\cdot\nabla\beta + {\bf \Gamma}\cdot\nabla\alpha\,\\
   \frac{\partial\dot{\beta}}{\partial x^3} & = -\nabla E_\parallel\cdot\nabla\alpha - {\bf \Gamma}\cdot\nabla\beta.
\end{align}
\end{subequations}
Because both $\nabla\alpha$ and $\nabla\beta$ are perpendicular to $\hat{\bf e}_3$, only the perpendicular components of $\nabla E_\parallel$ and $\bf \Gamma$, namely, $\nabla_\perp E_\parallel$ and $\bf \Gamma_\perp$, have effects on the local line conservation.
Considering that $\nabla_\perp E_\parallel$ and $\bf \Gamma_\perp$ can be explicitly calculated in the $\left(\hat{\bf e}_1,\hat{\bf e}_2,\hat{\bf e}_3\right)$ frame, we can use them to qualitatively compare $E_\parallel$ and $\bf N_\perp$ terms even without knowing $\nabla\alpha$ and $\nabla\beta$.
However, because an absolute comparison still relies on the flux coordinate frame, $\nabla_\perp E_\parallel$ and $\bf \Gamma_\perp$ only provide a statistical perspective of approximate comparison.

\begin{figure*}
\centering
\includegraphics[width=0.6\textwidth]{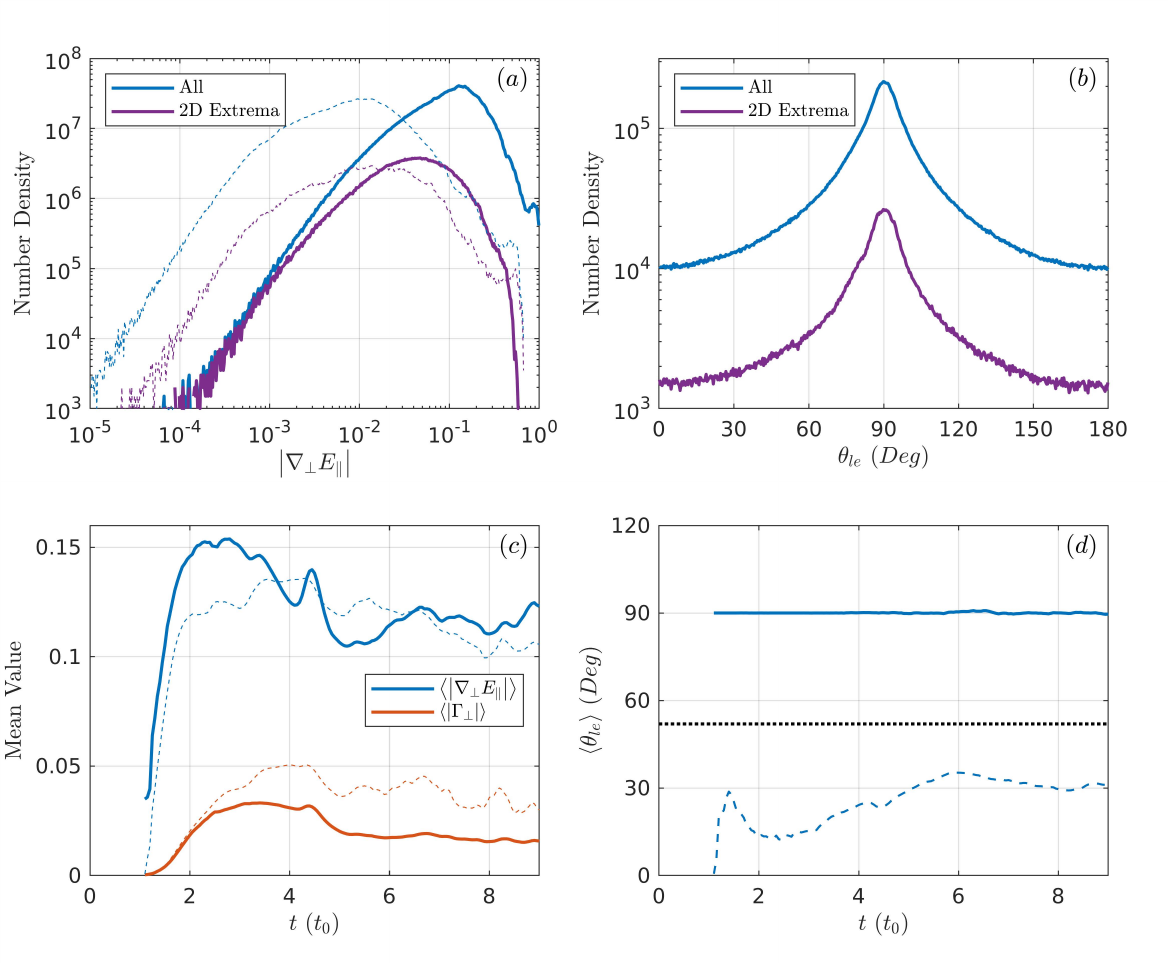}
\caption{Statistical comparisons of $\nabla_\perp E_\parallel$ and $\bf \Gamma_\perp$ in the Harris-sheet simulation.
Panel (a) plots the number densities of $\left|\nabla_\perp E_\parallel\right|$ (solid curves) and $\left|\bf \Gamma_\perp\right|$ (dashed curves) at $t=9$.
Panel (b) shows the number densities of the angle $\theta_{le}$ spanned by $\bf \Gamma_\perp$ and $\nabla_\perp E_\parallel$ at $t=9$.
In panels (a) and (b), the blue curves depict all grids satisfying $\left|J_\parallel\right|\geq 50$ while the purple ones plot the subsets with 2D extremal $E_\parallel$.
Panel (c) depicts the time evolutions of the mean values of $\left|\nabla_\perp E_\parallel\right|$ (the blue solid curve) and $\left|\bf \Gamma_\perp\right|$ (the orange solid curve).
The blue and orange dashed curves plot the standard deviations of $\left|\nabla_\perp E_\parallel\right|$ and $\left|\bf \Gamma_\perp\right|$, respectively.
Panel (d) exhibits the time evolutions of the mean values of $\theta_{le}$ (the solid curve) and its standard deviation (the dashed curve).
The black dashed line denotes the theoretical standard deviation of a uniform random distribution sampled from $0^\circ$ to $180^\circ$.
The mean values and standard deviations in panels (c) and (d) are calculated from all grids satisfying $\left|J_\parallel\right|\geq 50$.
}
\label{fig:localeffects}
\end{figure*}

For instance, we compare $\nabla_\perp E_\parallel$ and $\bf \Gamma_\perp$ in the Harris-sheet simulation shown in Sect.\,\ref{sec:test1} (see Fig.\,\ref{fig:localeffects}).
At $t=9$, the peak of $\left|\nabla_\perp E_\parallel\right|$ locates near $0.1$ while $\left|\bf \Gamma_\perp\right|$ mainly distributes near $0.01$ (see the blue curves in Fig.\,\ref{fig:localeffects}a).
Moreover, during the entire evolution, both the mean values and standard deviations of $\left|\nabla_\perp E_\parallel\right|$ are significantly larger than that of $\bf \Gamma_\perp$ (see Fig.\,\ref{fig:localeffects}c).
These results show that, statistically speaking, the magnitude of $\nabla_\perp E_\parallel$ is larger than $\bf \Gamma_\perp$.
On the other hand, we depict the number density of the angle $\theta_{le}$ spanned by $\nabla_\perp E_\parallel$ and $\bf \Gamma_\perp$, which shows that $\theta_{le}$ is most likely to be $90^\circ$, namely, the two vectors tend to normal to each other for most of grids (see Fig.\,\ref{fig:localeffects}b).
During the evolution, the mean value of $\theta_{le}$ keeps close to $90^\circ$ and its standard deviation is smaller than that of a uniform distribution, which implies that the distributions of $\theta_{le}$ tend to concentrate near $90^\circ$.

It should also be noticed that the grids with extremal $E_\parallel$ identified by the algorithm introduced in Sect.\,\ref{sec:method} compose a subset about $12\%$ of all grids and have smaller values of $\left|\nabla_\perp E_\parallel\right|$ (see the purple solid curve in Fig.\,\ref{fig:localeffects}a).
Therefore, the algorithm of locating 2D extremal $E_\parallel$ is a good numerical approximation of $\nabla_\perp E_\parallel=0$.

When using the \texttt{ARD} code, users can enable the calculation of $\nabla_\perp E_\parallel$ and $\bf \Gamma_\perp$ by setting \texttt{ARD\_AnalyzeLocalEffects} as 1, which costs more RAM and computation resources.
All the derivatives in the expressions of $\nabla_\perp E_\parallel$ and $\bf \Gamma_\perp$ are first calculated in the original Cartesian frame by the 2-order central difference method and then transformed into the local frame by the transformation matrix $\mathcal{T}^{lmn}$ in Eq.\,\ref{eq:Tlocal}.
\texttt{ARD} also outputs $\left|\nabla\times\mathbf{N}\right|$ and $\left|\bf B\times\left(\nabla\times N\right)\right|$ to study the local flux and line conservations (see Table \ref{tab:lordextradata}).

\section{Examples of Magnetic Structures on the MPP\label{asec:diagram}}

\begin{figure*}
\centering
\includegraphics[width=1\textwidth]{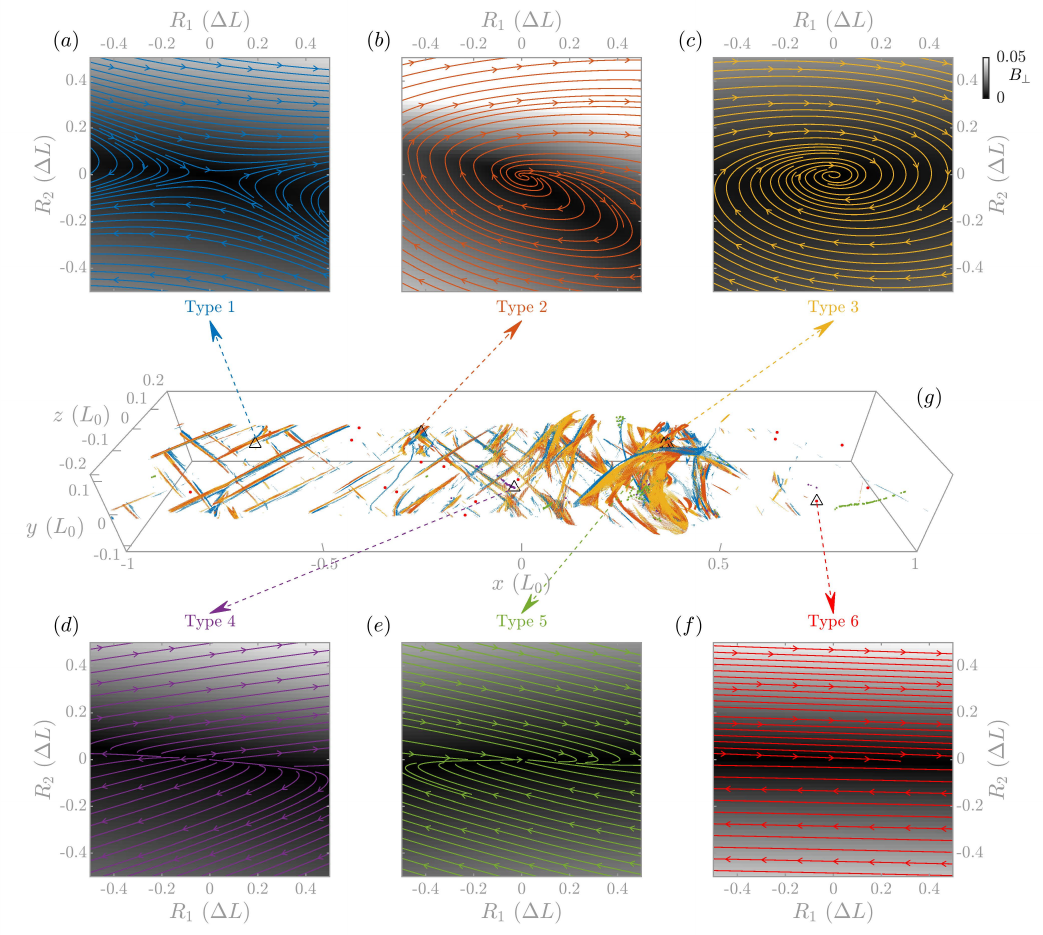}
\caption{Six typical local field-line structures of $\mathbf{B}_\perp$ (types 1 to 6) as found in the reconnection region at $t=9$.
Panels (a)--(f) depict the field lines of $\mathbf{B}_\perp$ on the MPP of selected grid positions as labeled by black triangle markers in panel (g).
Different colors correspond to different types.
The distributions of $B_\perp=\left|\mathbf{B}_\perp\right|$ are also plotted with a gray color map.
In panel (g), to give a clear picture, we only plot the grids with $\theta_\mathrm{eig}$ larger than a threshold value.
For types 1--3, the threshold is set as $30^\circ$; for types 4--5, the threshold is $5^\circ$.
}
\label{fig:diagram}
\end{figure*}

In Fig.\,\ref{fig:diagram}, we exhibit the $\mathbf{B}_\perp$ lines of types 1--6 as found in the simulation of 3D Harris-sheet reconnection, which proves that our method can precisely capture the local magnetic structures.

\section{Quantitative Statistical Results of O-Type Grids in the Harris-sheet Simulation\label{asec:OGrids}}

\begin{figure*}
\centering
\includegraphics[width=0.6\textwidth]{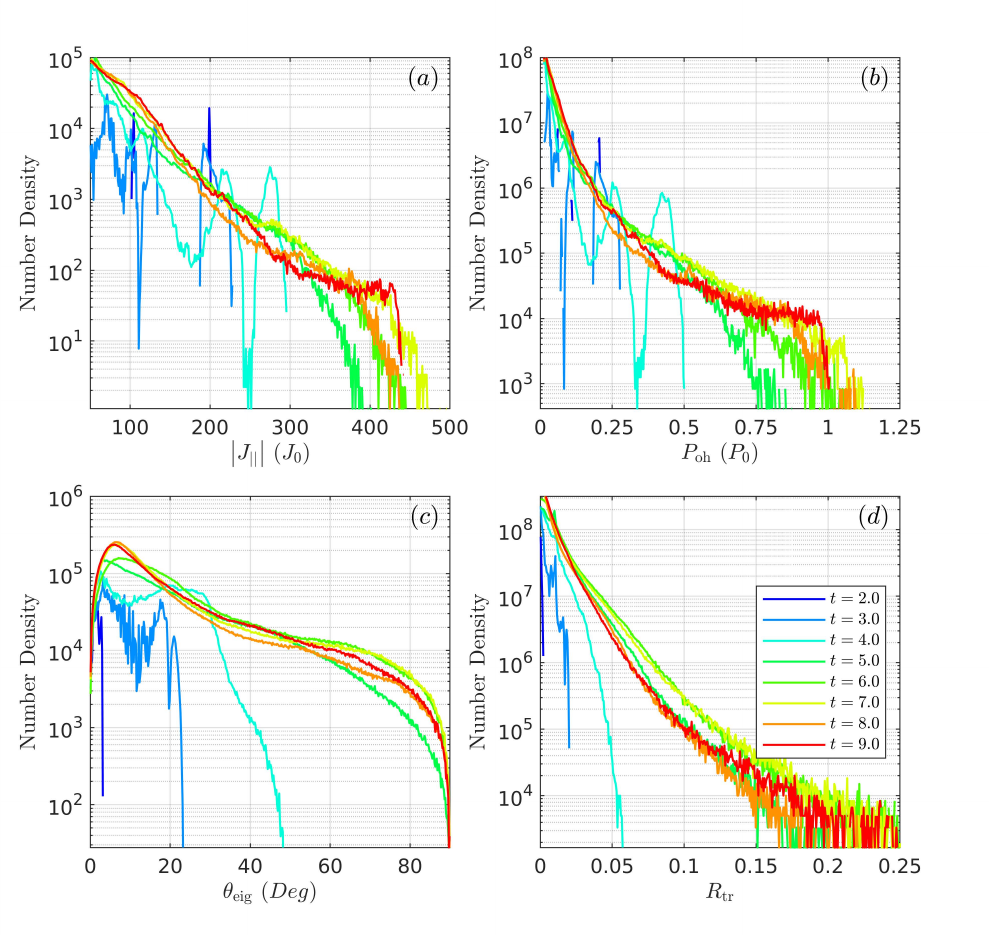}
\caption{The number densities of $\left|J_\parallel\right|$ (a) , $P_\mathrm{oh}$ (b), $\theta_\mathrm{eig}$ (c), and $R_\mathrm{tr}$ (d) of the 3D O-type grids at different moments.
Different colors correspond to different moments.
}
\label{fig:hist_o_alltime}
\end{figure*}

\begin{figure*}
\centering
\includegraphics[width=0.6\textwidth]{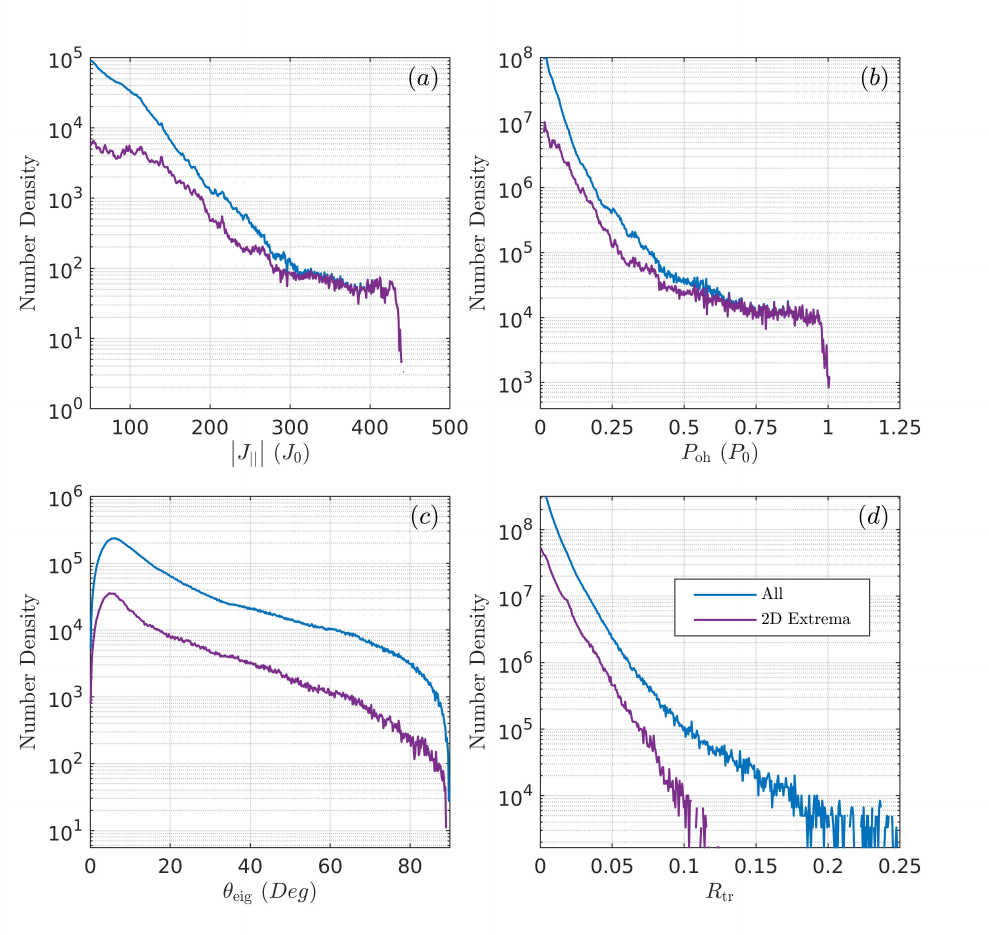}
\caption{The number densities of $\left|J_\parallel\right|$ (a), $P_\mathrm{oh}$ (b), $\theta_\mathrm{eig}$ (c), and $R_\mathrm{tr}$ (d) of the 3D O-type grids at $t=9$.
The blue curves depict profiles of all 3D O-types grids, while the purple ones plot the subsets with 2D extrema of $J_\parallel$.
}
\label{fig:hist_o_180}
\end{figure*}

Figures \ref{fig:hist_o_alltime} and \ref{fig:hist_o_180} plot the same statistical results as Figs.\,\ref{fig:hist_x_alltime} and \ref{fig:hist_x_180} but for 3D O-type grids.

\end{appendix}

\end{document}